\documentclass[aps,prb,twocolumn,superscriptaddress,amsmath,amssymb,longbibliography]{revtex4-2}

\usepackage{graphicx}
\usepackage{dcolumn}
\usepackage{bm}
\usepackage{subfig}
\begin{document}
	
	\title{Nearest-Neighbor Tight-Binding Realization of Hyperbolic Lattices with $\mathbb{Z}_2$ Gauge Structures}
	
	
	\author{Xianghong Kong}
	\affiliation{Department of Electrical and Computer Engineering, National University of Singapore, Singapore, Singapore}
	\affiliation{National University of Singapore Suzhou Research Institute, No. 377 Linquan Street, Suzhou, Jiangsu, China}
	\author{Xingsi Liu}
	\affiliation{Department of Electrical and Computer Engineering, National University of Singapore, Singapore, Singapore}
	\author{Shuihua Yang}
	\affiliation{Department of Electrical and Computer Engineering, National University of Singapore, Singapore, Singapore}
	\author{Zhiyuan Yan}
	\affiliation{Department of Electrical and Computer Engineering,
		Faculty of Science and Technology,
		University of Macau, Macau, SAR, China}

	\author{Weijin Chen}
	\affiliation{Institute of Precision Optical Engineering, School of Physics Science and Engineering, Tongji University, Shanghai, China}
	\author{Zhixia Xu}
	\email{zhixiaxu@seu.edu.cn}
	\affiliation{State Key Laboratory of Millimeter Waves, Southeast University, Nanjing 210096, China}
	
	\author{Cheng-Wei Qiu}
	\email{chengwei.qiu@nus.edu.sg}
	\affiliation{Department of Electrical and Computer Engineering, National University of Singapore, Singapore, Singapore}
    \affiliation{National University of Singapore Suzhou Research Institute, No. 377 Linquan Street, Suzhou, Jiangsu, China}
	
	\date{\today}
\begin{abstract}
A systematic framework for realizing $\mathbb{Z}_2$ gauge extensions of hyperbolic lattices within the nearest-neighbor tight-binding formalism is developed. Using the triangle group $\Delta(2,8,8)$ as an example, we classify all inequivalent projective symmetry groups by computing the second cohomology group $H^2(\Delta(2,8,8),\mathbb{Z}_2)$. Each class corresponds to a distinct flux configuration and can be constructed by tight-binding models to verify the symmetry relations of the extended group. The translation subgroups of the $\mathbb{Z}_2$ extended lattices are associated with high genus surfaces, which follows the Riemann-Hurwitz formula. By applying the Abelian hyperbolic band theory, we find the all-flat dispersions along specific directions in momentum space and van Hove singularities correlated with discrete eigenenergies. Our results establish a general route to investigate gauge-extended hyperbolic lattices and provide a foundation for further studying symmetry fractionalization and spin liquid phases in non-Euclidean geometries.
\end{abstract}
	
	
	\maketitle
	\section{Introduction}
While Euclidean lattices have long served as the foundation for condensed matter physics, recent advances have extended the study to hyperbolic lattices which are periodic in two-dimensional hyperbolic space. The	Schl\"afli symbol \{$p,q$\} is applied to describe the regular tessellations where $p$ denotes the number of edges of the regular polygon and $q$ denotes the number of polygons meeting at each vertex. The expression $(p-2)(q-2)>4$ is satisfied for the hyperbolic lattices. One critical breakthrough in this research area is the generalization of Bloch theory in Euclidean space, which leads to the Abelian hyperbolic band theory \cite{maciejko2021hyperbolic, boettcher2022crystallography,maciejko2022automorphic}. 
The discovery of hyperbolic topological band insulators \cite{urwyler2022hyperbolic,zhang2023hyperbolic,sun2024topological}, hyperbolic flat bands \cite{guan2025topological,yuan2024hyperbolic} are enabled by the generalized band theory due to the characterization of the spectra in momentum space. However, while Euclidean translation groups are Abelian, the hyperbolic translation groups are non-Abelian and require higher-dimensional irreducible representations (IRs) besides the Abelian IRs. Different approaches such as the coherent sequence of normal subgroups method \cite{lenggenhager2023non,lux2023converging}, and the continued-fraction method \cite{mosseri2023density} have been developed to construct non-Abelian hyperbolic band theory.

	\begin{figure}
	\centering
	\vspace{-0.4cm}
	\subfloat[]{{\includegraphics[width=.45\textwidth]{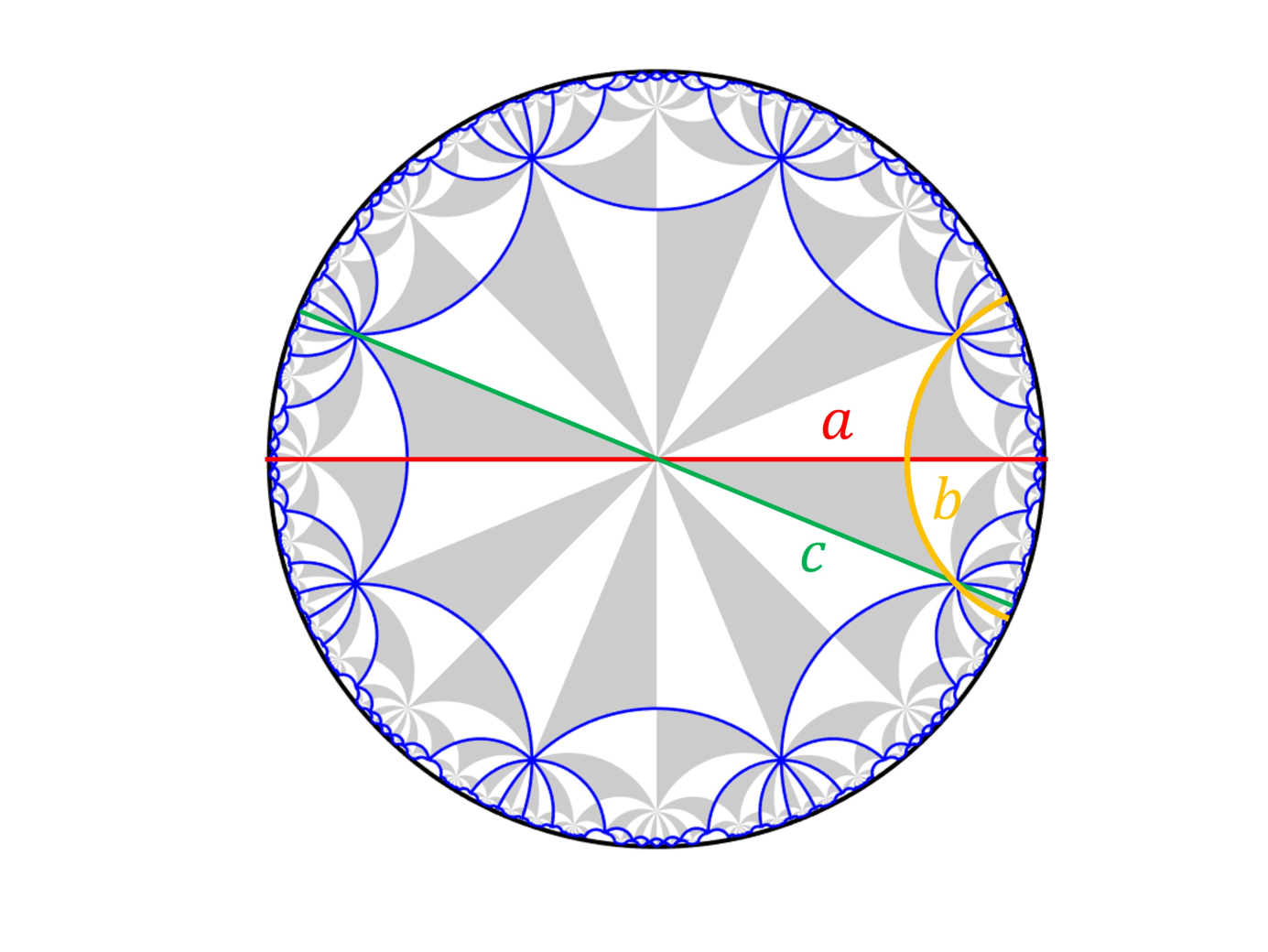}}\label{fig:hyper_symmetry}}\\
	\vspace{-0.4cm}
	\subfloat[]{{\includegraphics[width=.45\textwidth]{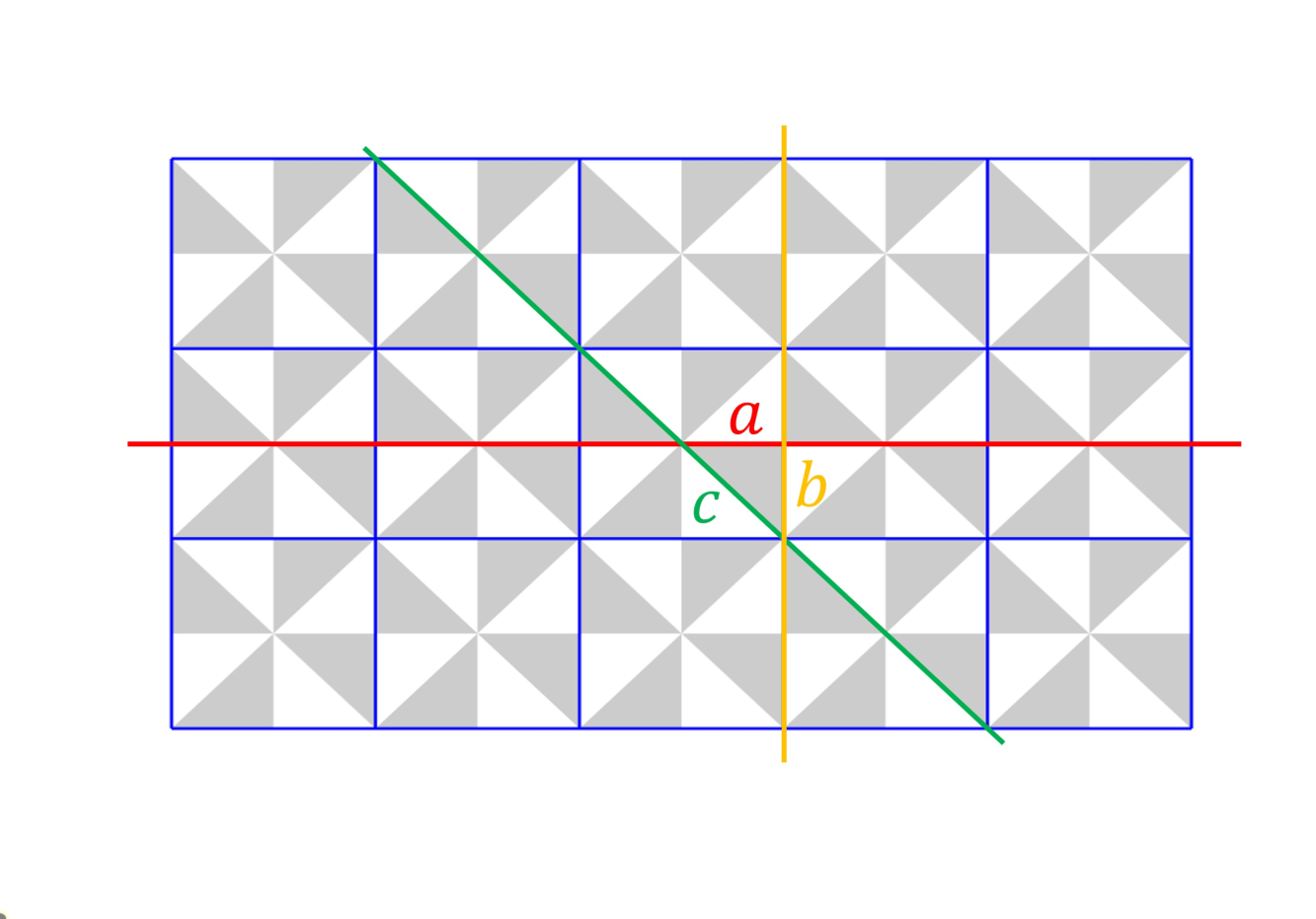}}\label{fig:square_symmetry}}\\
	\caption{Symmetries of different triangle groups. The primitive cells are plotted by blue polygons.(a) Triangle group $\Delta(2,8,8)$ presented by alternating grey and white triangles. The reflection lines $a, b, c$ are the generators shown in Eq.~(\ref{eq:triangle}). (b) Triangle group $\Delta(2,4,4)$ (square lattice) with its reflection lines $a, b, c$. }
	\label{fig:symmetry} 
\end{figure}

Symmetry has been the most fundamental concept in the exploration of condensed matter physics \cite{el2008symmetry,hu2023source}. Embedding gauge structures into lattice models usually will reduce the symmetry groups of the systems \cite{li2022transition,koga2021majorana}. However, Wen has introduced the projective symmetry group on the square lattice which can still preserve all the original symmetries up to a gauge transformation \cite{wen2002quantum}. This framework can also be applied to chiral spin liquids on triangular and kagome lattices \cite{bieri2016projective}, and other wallpaper groups in two dimensions \cite{chen2023classification}. Recently,  a $\mathbb{Z}_2$  gauge field was introduced to the hyperbolic lattice to investigate the hyperbolic spin liquid and chiral quasiparticles with non-Abelian Bloch profile \cite{dusel2025chiral,lenggenhager2025hyperbolic}. However, a thorough projective symmetry group classification of the  $\mathbb{Z}_2$  gauge extension on a hyperbolic lattice has not been studied yet.

In this work, we develop the projective symmetry group classification of $\mathbb{Z}_2$ extension on a hyperbolic lattice which is described by the triangle group $\Delta(2,8,8)$.  The second cohomology group $H^2(\Delta(2,8,8),\mathbb{Z}_2)$ is applied to classify inequivalent projective symmetry groups mathematically. Then simple nearest-neighbor tight-binding models are designed to realize all the classifications with different gauge fluxes.  Also, the Abelian hyperbolic band theory is applied to the tight-binding models to explore the band diagrams of hyperbolic lattices with $\mathbb{Z}_2$ gauge structures. The all-flat bands along a single specified line in momentum space are presented.
	
\section{\label{}Triangle group of the hyperbolic plane }
As shown in Fig. \ref{fig:hyper_symmetry}, the symmetries of a \{$p,q$\} hyperbolic lattice are described by the triangle group \cite{lenggenhager2023non}
		\begin{equation}
	\Delta(r,q,p)=\langle a,b,c|a^2,b^2,c^2,(ab)^r,(bc)^q,(ca)^p\rangle,
		\label{eq:triangle}
	\end{equation}
where $a,b,c$ are reflections over different sides of a triangle with internal angles $\pi/r$, $\pi/q$, and $\pi/p$. Assume the triangle surrounded by three reflection lines is denoted as the identity of the triangle group, all the triangles shown in Fig. \ref{fig:hyper_symmetry} can be represented by the word built from the generators $a$,$b$, and $c$. We use the right-action convention for convenience, where the word $ab$ means "apply reflection $a$ then $b$". For comparison, the wallpaper group p4m is also written in the form of the triangle group $\Delta(2,4,4)$, which can be verified by Fig. \ref{fig:square_symmetry}.
\begin{figure}
		\centering
		
		\subfloat[]{{\includegraphics[width=.45\textwidth]{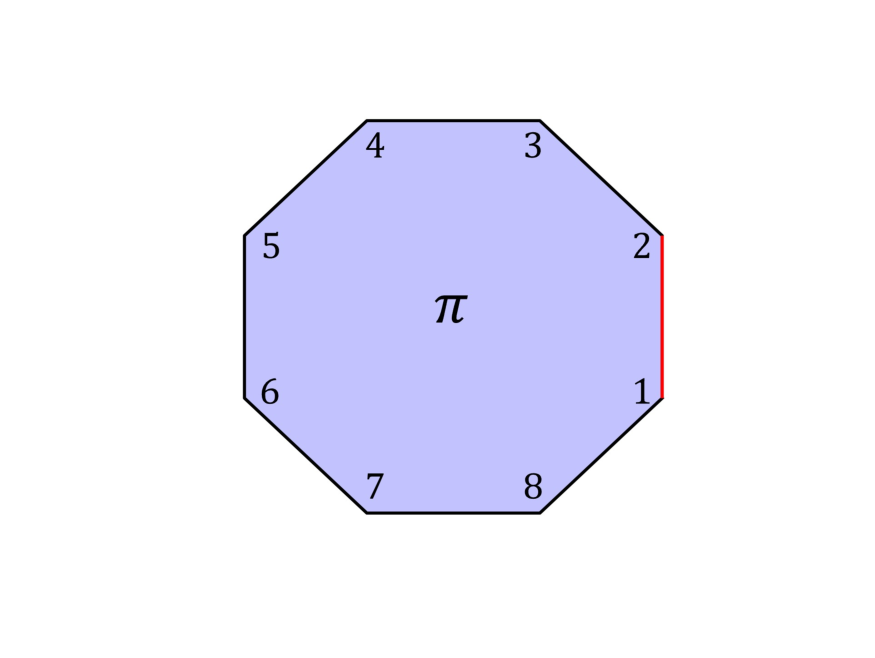}}\label{fig:flux_rotation}}
		\\
		\vspace{-0.6cm}
		\subfloat[]{{\includegraphics[width=.3\textwidth]{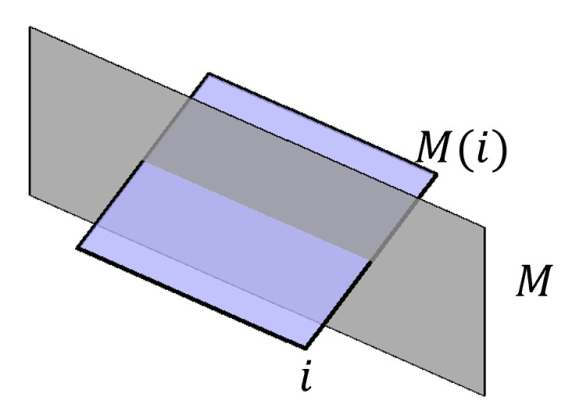}}\label{fig:flux_mirror1}}
		\hspace{-0.6cm}
		\subfloat[]{{\includegraphics[width=.20\textwidth]{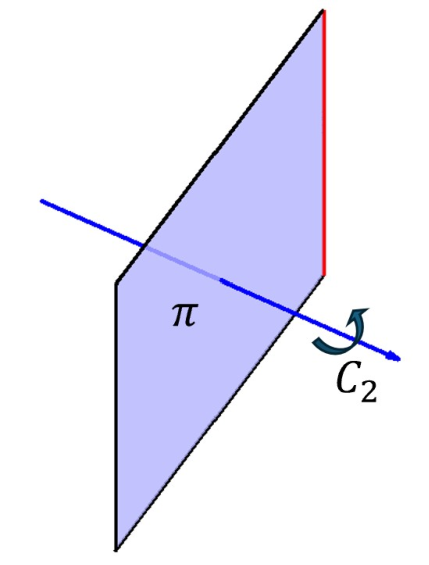}}\label{fig:flux_mirror2}}
		\caption{(a) The $\pi$-flux of rotation operator $\tilde{R}^4=-1$. (b) Conventional mirror symmetry $M$ is replaced by (c) a $C_2$ rotation around a horizontal axis. Red edges represent hopping coefficients $t_{ij}=-1$ while black edges represent $t_{ij}=1$. }
		\label{fig:flux} 
\end{figure}

\section{Projective representations of the triangle group $\Delta(2,8,8)$}

A projective representation of $G$ with coefficients in an Abelian group $A$ is a map
\begin{equation}
	\tilde{\rho}: G \rightarrow GL(V)
\end{equation} 
such that a modified multiplication rule is satisfied:
\begin{equation}
	\tilde{\rho}(g_1)\tilde{\rho}(g_2)=\omega(g_1,g_2)\tilde{\rho}(g_1g_2),\quad  \forall g_1,g_2\in G.
	\label{eq:modified_multiplication}
\end{equation}
Mathematically, all the inequivalent factor systems $\omega(g_1,g_2), \forall g_1,g_2\in G$ form the second cohomology group $H^2(G,A)$ (see Appendix~\ref{appendix:cohomology}). In our case, in order to find out all the projective symmetry groups of the $\mathbb{Z}_2$  gauge extension on the hyperbolic lattice $\Delta(2,8,8)$, we need to figure out $H^2(\Delta(2,8,8),\mathbb{Z}_2)$. Each equivalent factor $[\omega]\in H^2(\Delta(2,8,8),\mathbb{Z}_2)$  corresponds to a distinct central extension which can be expressed as:
		\begin{equation}
	1 \rightarrow \mathbb{Z}_2 \rightarrow \tilde{\Delta}(2,8,8) \rightarrow \Delta(2,8,8) \rightarrow 1.
	\label{eq:exact_sequence}
\end{equation}
Since the triangle group $\Delta(2,8,8)$ has infinite group elements, it is hard to list the factor system $\omega(g_1,g_2), \forall g_1,g_2\in \Delta(2,8,8)$ of the projective symmetry group $\tilde{\Delta}(2,8,8)$. Instead, we only need to take the relators of Eq.~(\ref{eq:triangle}) into consideration when deriving the $H^2(\Delta(2,8,8),\mathbb{Z}_2)$. Following the method used in \cite{chen2023classification}, we can express the projective relations of the triangle group $\Delta(2,8,8)$ as
\begin{subequations}
	\begin{eqnarray}
		\tilde{a}^2=\beta_1,
		\label{eq:288relation1}
	\end{eqnarray}
	\begin{eqnarray}
		\tilde{b}^2=\beta_2,
		\label{eq:288realtion2}
	\end{eqnarray}
	\begin{eqnarray}
		\tilde{c}^2=\beta_3,
		\label{eq:288realtion3}
	\end{eqnarray}
	\begin{eqnarray}
		(\tilde{a}\tilde{b})^2=\alpha_1,
		\label{eq:288realtion4}
	\end{eqnarray}
	\begin{eqnarray}
		(\tilde{b}\tilde{c})^8=\alpha_2,
		\label{eq:288realtion5}
	\end{eqnarray}
	\begin{eqnarray}
		(\tilde{c}\tilde{a})^8=\alpha_3,
		\label{eq:288realtion6}
	\end{eqnarray}
	\label{eq:288relation}
\end{subequations}
where the six independent cohomology invariants $\beta_1, \beta_2, \beta_3, \alpha_1, \alpha_2, \alpha_3\in \mathbb{Z}_2$. By redefining the generator such as $\tilde{a}\prime=-\tilde{a}$ won't change the projective relations shown in Eq.~(\ref{eq:288relation}) due to the even indices of the expressions. Hence we have
\begin{equation}
	H^2(\Delta(2,8,8),\mathbb{Z}_2)=\mathbb{Z}_2^6.
	\label{eq:H2ofhyperbolic}
\end{equation}
The result can also be verified easily by the HAP package \cite{HAP} of the computational algebra system GAP. For comparison, the wallpaper group p4m which can be described as the triangle group $\Delta(2,4,4)$ has $H^2(\Delta(2,4,4),\mathbb{Z}_2)=\mathbb{Z}_2^6$ \cite{chen2023classification}.

\section{Tight-binding models with respective cohomology invariants}
\begin{figure}
	\centering
	\hspace*{-0.5cm}
	\subfloat[]{{\includegraphics[width=.53\textwidth]{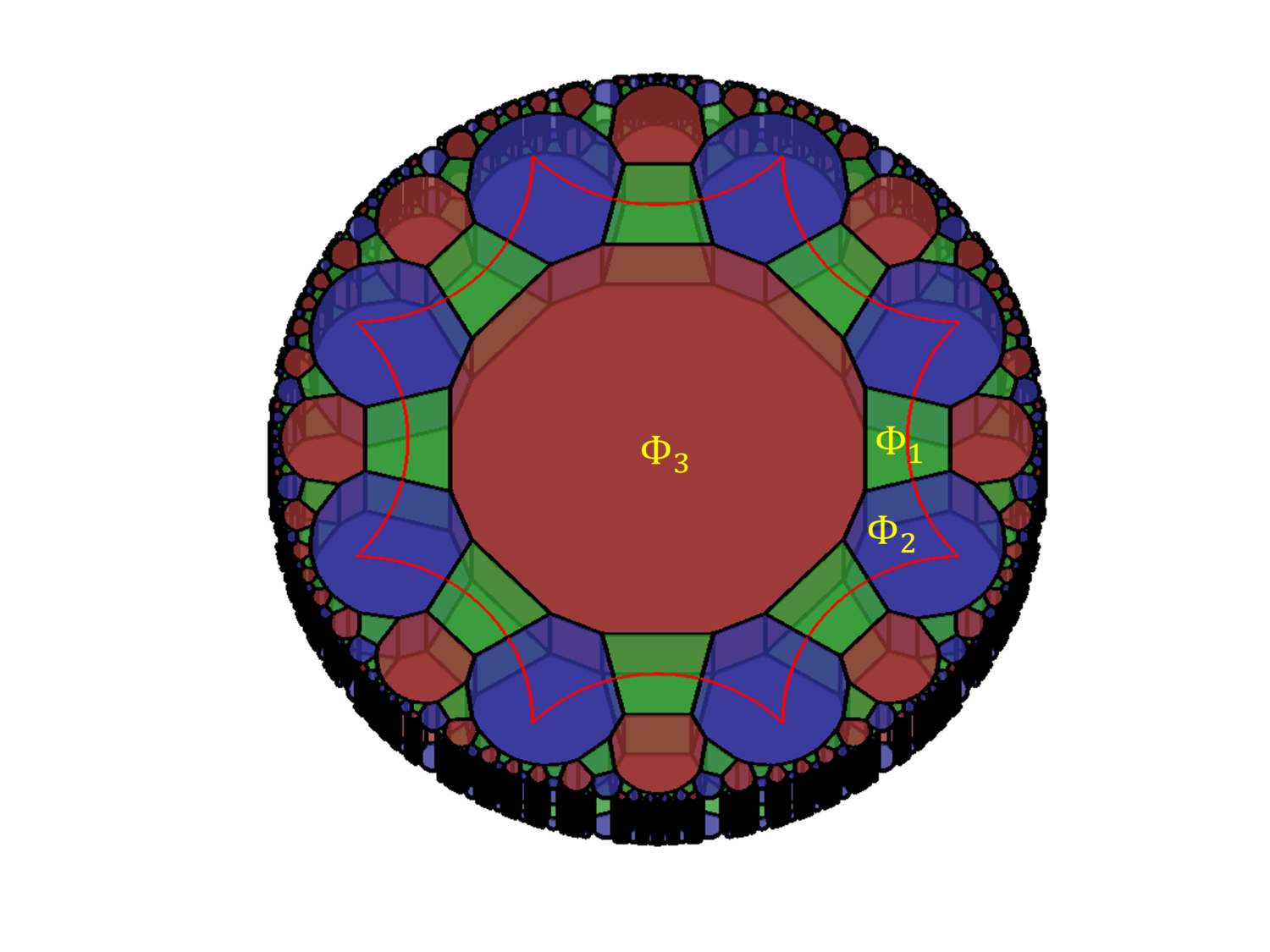}}\label{fig:hyper_flux1}}
	\\
	\hspace*{-0.5cm}
	\subfloat[]{{\includegraphics[width=.53\textwidth]{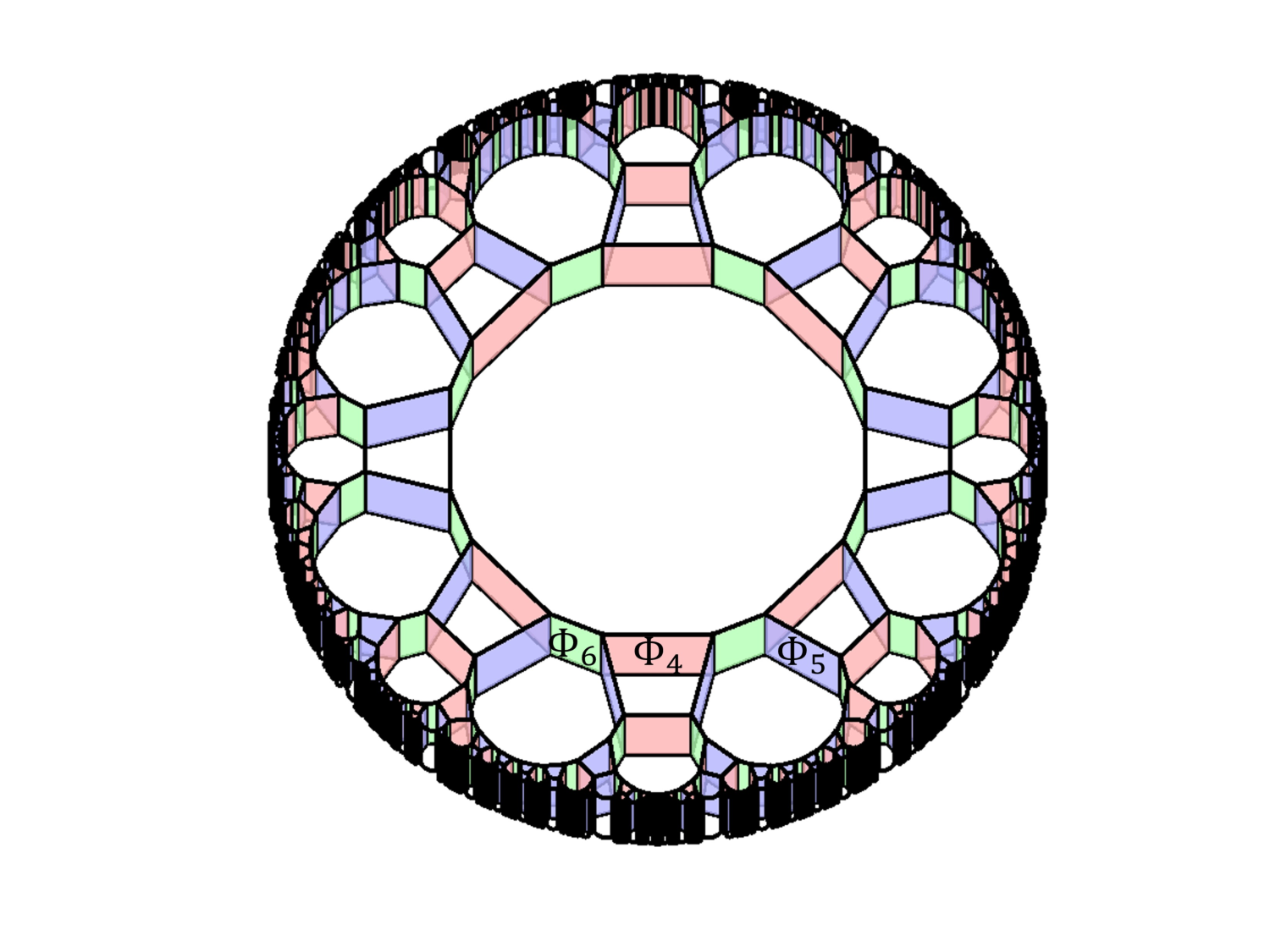}}\label{fig:hyper_flux2}}\\
	\caption{Flux distributions  with different cohomology invariants of the triangle group $\Delta(2,8,8)$.(a) The fluxes $\Phi_i, i=1, 2, 3$ correspond to cohomology  invariants $\alpha_i, i=1, 2, 3$ as shown in Eq.~(\ref{eq:288relation}). (b) The fluxes $\Phi_i, i=4, 5, 6$ correspond to cohomology  invariants $\beta_i, i=1, 2, 3$ as shown in Eq.~(\ref{eq:288relation}). }
	\label{fig:hyper_flux} 
\end{figure}
\subsection{fluxes with different cohomology invariants}
The 6 relators shown in Eq.~(\ref{eq:288relation}) can be classified into two categories: relators of the mirror symmetry operators and relators of the rotation symmetry operators. For the discrete lattice model, the mirror symmetry operator $M$ and the rotation symmetry operator $R$ can be viewed as the permutation matrices while the corresponding gauge transformations $G_M$ and $G_R$ can be presented by diagonal matrices with values $\pm 1$ for different lattice sites. The relations between projective representations and  spatial symmetry representations without a gauge field can be expressed as \cite{chen2023classification}
\begin{subequations}
	\begin{eqnarray}
		\tilde{R}=G_RR,
		\label{eq:relationrr}
	\end{eqnarray}
	\begin{eqnarray}
		\tilde{M}=G_MM.
		\label{eq:relationmm}
	\end{eqnarray}
\end{subequations}
\subsubsection{cohomology invariants of rotation operators}
Assume the tight-binding model is written as $H=\sum_{ij}t_{ij}a_i^{\dagger}a_j$. After applying the rotation operator $\tilde{R}$ to $H$, we have
\begin{equation}
	H^{\prime}=\sum_{ij}G_R(i)G_R^{-1}(j)t_{R^{-1}(i)R^{-1}(j)}a_i^{\dagger}a_j,
\end{equation} 
where $G_R(i)$ is the gauge transformation of the rotation operator on site i. Since the system is invariant under the transformation $\tilde{R}$, we have 
\begin{equation}
	t_{ij}=G_R(i)G_R^{-1}(j)t_{R^{-1}(i)R^{-1}(j)}.
	\label{eq:coefficient}
\end{equation}
Since nontrivial cohomology invariant $\tilde{R}^n=\alpha_r$ only appears when $n$ is even, we can define the two-fold rotation $\tilde{R}_{\pi}^2$ by
\begin{equation}
	\tilde{R}_{\pi}^2=\tilde{R}^n=\alpha_r.
	\label{eq:def_R_pi}
\end{equation}
Combining Eq.~(\ref{eq:def_R_pi}) and Eq.~(\ref{eq:relationrr}), we can achieve 
\begin{eqnarray}
	\tilde{R}_{\pi}^2&=&G_{R_{\pi}}R_{\pi}G_{R_{\pi}}R_{\pi}=G_{R_{\pi}}R_{\pi}G_{R_{\pi}}R_{\pi}^{-1}R_{\pi}^2 \nonumber \\
	&=&G_{R_{\pi}}(\mathbf{r})G_{R_{\pi}}(R_{\pi}(\mathbf{r}))=\alpha_r.
	\label{eq:rotation_cohomology}
\end{eqnarray}

For a discrete lattice model, we need an even number of sites $i=1, 2, \cdots 2l$ to represent a system which is symmetric under $R_n$ ($n$ is even). As shown in Fig. \ref{fig:flux_rotation}, the relations between hopping coefficients $t_{i,i+1}$ and $t_{i+l,i+l+1}$ can be derived from Eq.~(\ref{eq:coefficient}) by replacing operator $R$ with $R_{\pi}$:
\begin{equation}
	t_{i,i+1}=G_{R_{\pi}}(i)G_{R_{\pi}}^{-1}(i+1)t_{i+l,i+l+1}.
	\label{eq:pi_rotation}
\end{equation}
Following the convention in ref.~\cite{chen2023classification}, the flux of the area shown in Fig. \ref{fig:flux_rotation} can be expressed as
\begin{eqnarray}
	\exp(-i\Phi)&=&t_{2l,1}\prod_{i=1}^{2l-1}t_{i,i+1}\nonumber\\
	&=&t_{2l,1}\prod_{i=1}^{l}t_{i,i+1}\prod_{i=l+1}^{2l-1}t_{i,i+1}.
	\label{eq:flux_rotationstep1}
\end{eqnarray}
Here, we assume the hopping coefficients $t_{i,i+1}$ are normalized. Substituting Eq.~(\ref{eq:pi_rotation}) into Eq.~(\ref{eq:flux_rotationstep1}) for $t_{i,i+1}, i\leq l$, we get 
\begin{equation}
	\exp(-i\Phi)=G_{R_{\pi}}(1)G_{R_{\pi}}^{-1}(l+1)=\alpha_r,
	\label{eq:relation_rotationflux_cohomogy}
\end{equation}
where $t_{i,i+1}^2=1$ and $G_{R_{\pi}}^{2}(i)=1$ since we focus on the $\mathbb{Z}_2$ extension.

\subsubsection{cohomology invariants of mirror operators}
Similarly, for the mirror operator $\tilde{M}^2=\alpha_m$, Eq.~(\ref{eq:coefficient}) and Eq.~(\ref{eq:rotation_cohomology}) can be rewritten as:
\begin{equation}
	t_{M(i),i}=G_M(M(i))G_{M}^{-1}(i)t_{i,M(i)},
	\label{eq:hopping_mirror}
\end{equation}

\begin{eqnarray}
	\tilde{M}^2&=&G_MMG_MM=G_MMG_MM^{-1}M^2\nonumber\\
	&=&G_M(\mathbf{r})G_M(M(\mathbf{r}))=\alpha_m.
	\label{eq:mirror_cohomology}
\end{eqnarray}
Combing Eq.~(\ref{eq:mirror_cohomology}) with Eq.~(\ref{eq:hopping_mirror}) we can get
\begin{equation}
	t_{M(i),i}=\alpha_m t_{i,M(i)},
\end{equation}
where $G_{M}^2(i)=1$ is still valid. When $\alpha_m=-1$, we have $t_{M(i),i}=-t_{i,M(i)}$, which can not be realized by real nearest-neighbor tight-binding models (see Fig. \ref{fig:flux_mirror1}). Hence, a bilayer version of the lattice model is raised to replace the mirror operator $M$ with a $C_2$ rotation around a horizontal axis as shown in Fig. \ref{fig:flux_mirror2} \cite{chen2023classification}. Since we have derived the relation between the flux of the rotation operator and the cohomology invariant, the result in Eq.~(\ref{eq:relation_rotationflux_cohomogy}) is still valid for the $C_2$ rotation around the horizontal axis.

The flux distributions with different cohomology invariants of the triangle group $\Delta(2,8,8)$ are concluded in Fig. \ref{fig:hyper_flux}, where the three cohomology invariants of rotation operators are determined by the flux distributions of intralayer plaquettes and the mirror operators are determined by the flux distributions of interlayer plaquettes.

\begin{figure}
	\centering
	\includegraphics[width=.5\textwidth]{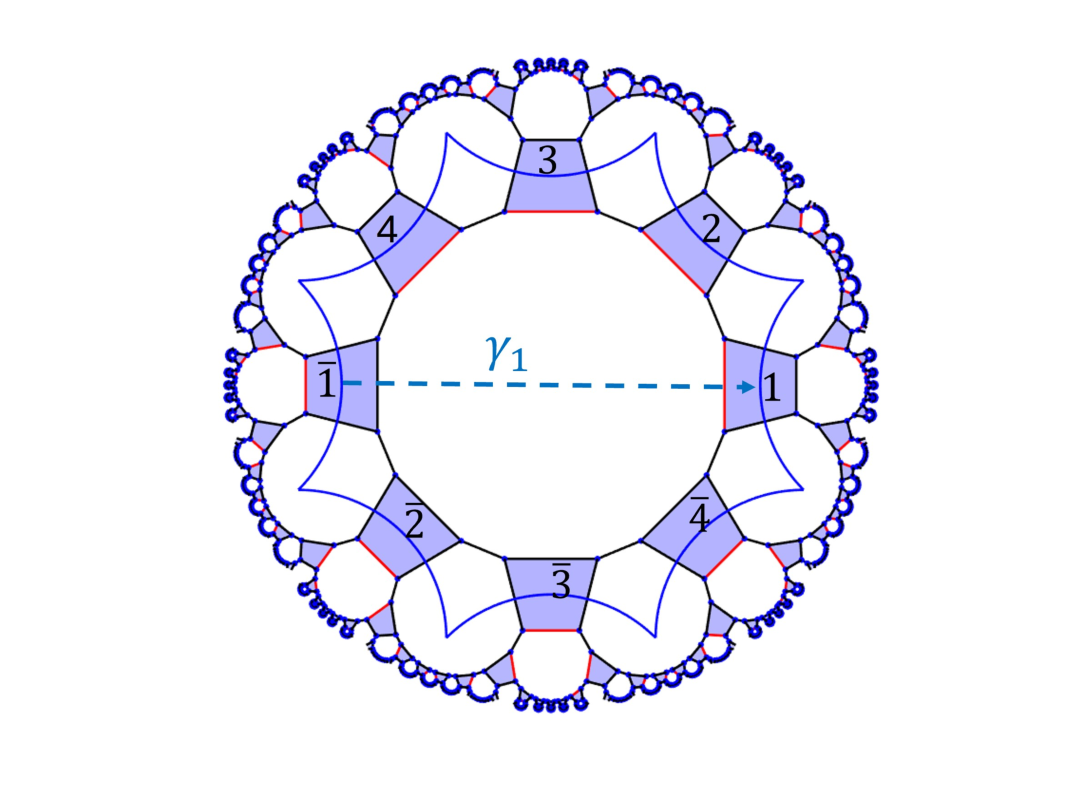}
	
	\caption{Top-down projection schematic of the tight-binding model with flux distribution $(\alpha_1, \alpha_2, \alpha_3, \beta_1, \beta_2, \beta_3)=(-1, 1, 1, 1, 1, 1)$. The black/red lines represent hopping coefficients $t_{ij}=1/-1$ for both top and bottom layers. The blue dots represent the hopping coefficients $t_{ij}=1$ between the top and bottom layers. The primitive cell is shown by blue curved lines and the periodic boundaries $\bar{i}$ and $i$ are related by the translation operator $\gamma_i$.}
	\label{fig:phi1} 
\end{figure}
	\begin{figure*}
	\centering
	
	\includegraphics[width=\textwidth]{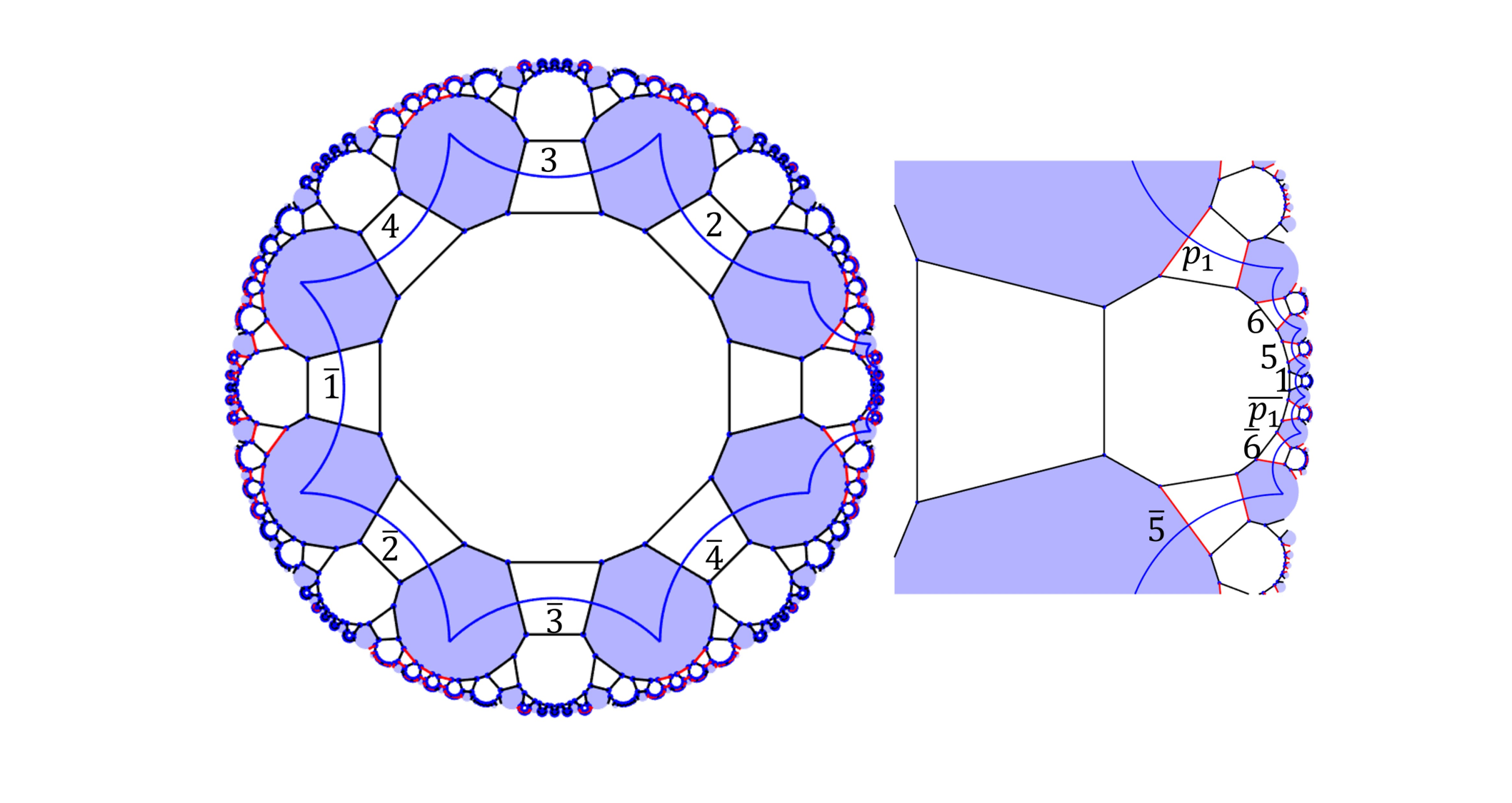}
	
	\caption{Top-down projection schematic of the tight-binding model with flux distribution $(\alpha_1, \alpha_2, \alpha_3, \beta_1, \beta_2, \beta_3)=(1, -1, 1, 1, 1, 1)$. The black/red lines represent hopping coefficients $t_{ij}=1/-1$ for both top and bottom layers. The blue dots represent the hopping coefficients $t_{ij}=1$ between the top and bottom layers. The supercell is shown by blue curved lines and the periodic boundaries $\bar{i}$ and $i$ are related by the translation operator $\tilde{\gamma}_i$.}
	\label{fig:phi2} 
\end{figure*}
	\begin{figure}
	\centering
	\includegraphics[width=0.55\textwidth]{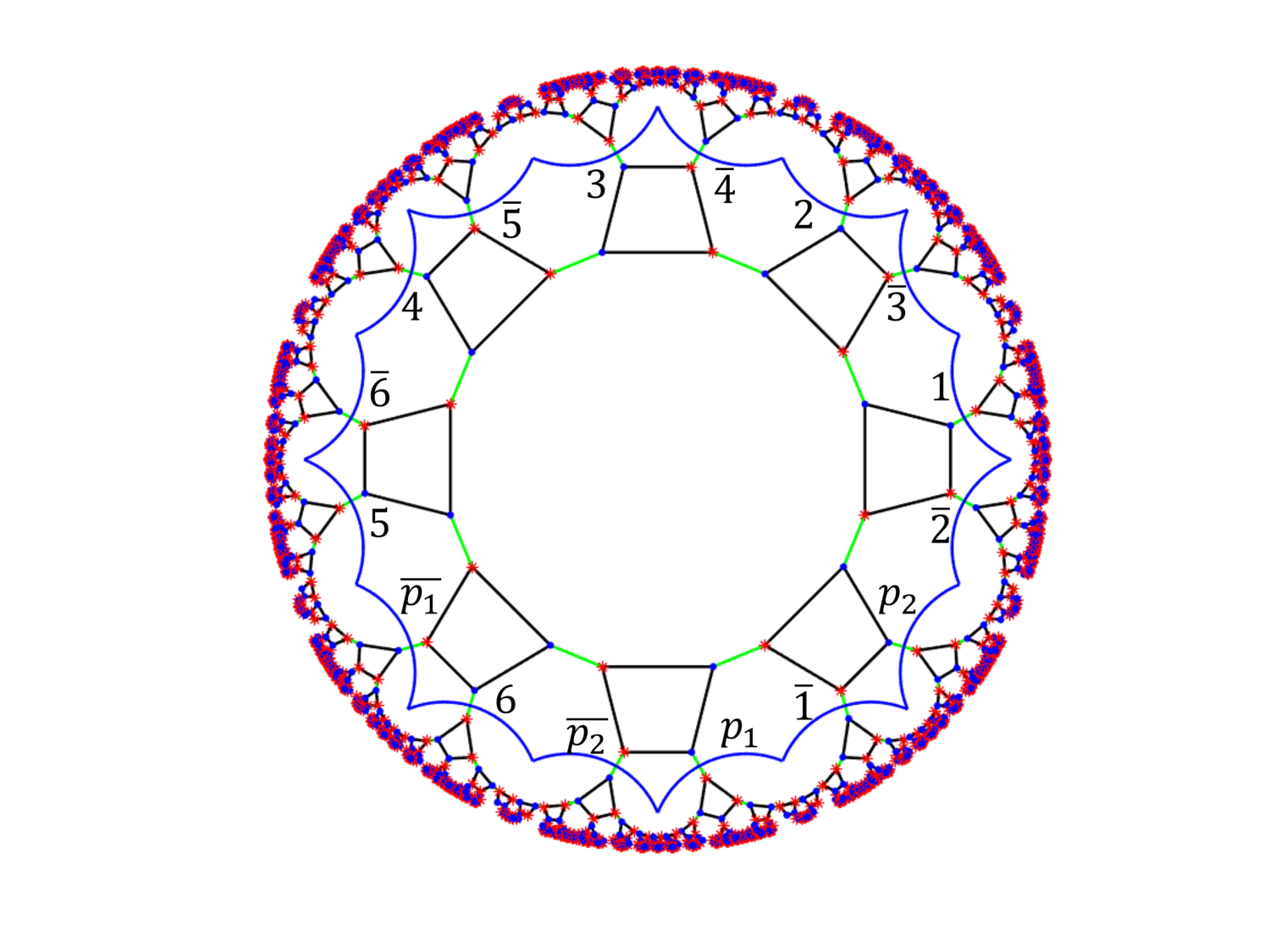}
	\caption{Top-down projection schematic of the tight-binding model with flux distribution $(\alpha_1, \alpha_2, \alpha_3, \beta_1, \beta_2, \beta_3)=(1, 1, 1, -1, 1, 1)$. The black lines represent hopping coefficients $t_{ij}=1$ for both top and bottom layers while the green lines represent hopping coefficients $t_{ij}=1$ for the top layer and $t_{ij}=-1$ for the bottom layer. The blue dots/red stars represent the hopping coefficients $t_{ij}=1/-1$ between the top and bottom layers. The supercell is shown by blue curved lines and the periodic boundaries $\bar{i}$ and $i$ are related by the translation operator $\tilde{\gamma}_i$.}
	\label{fig:phi4} 
\end{figure}
\begin{figure*}
	\centering
	
	\includegraphics[width=\textwidth]{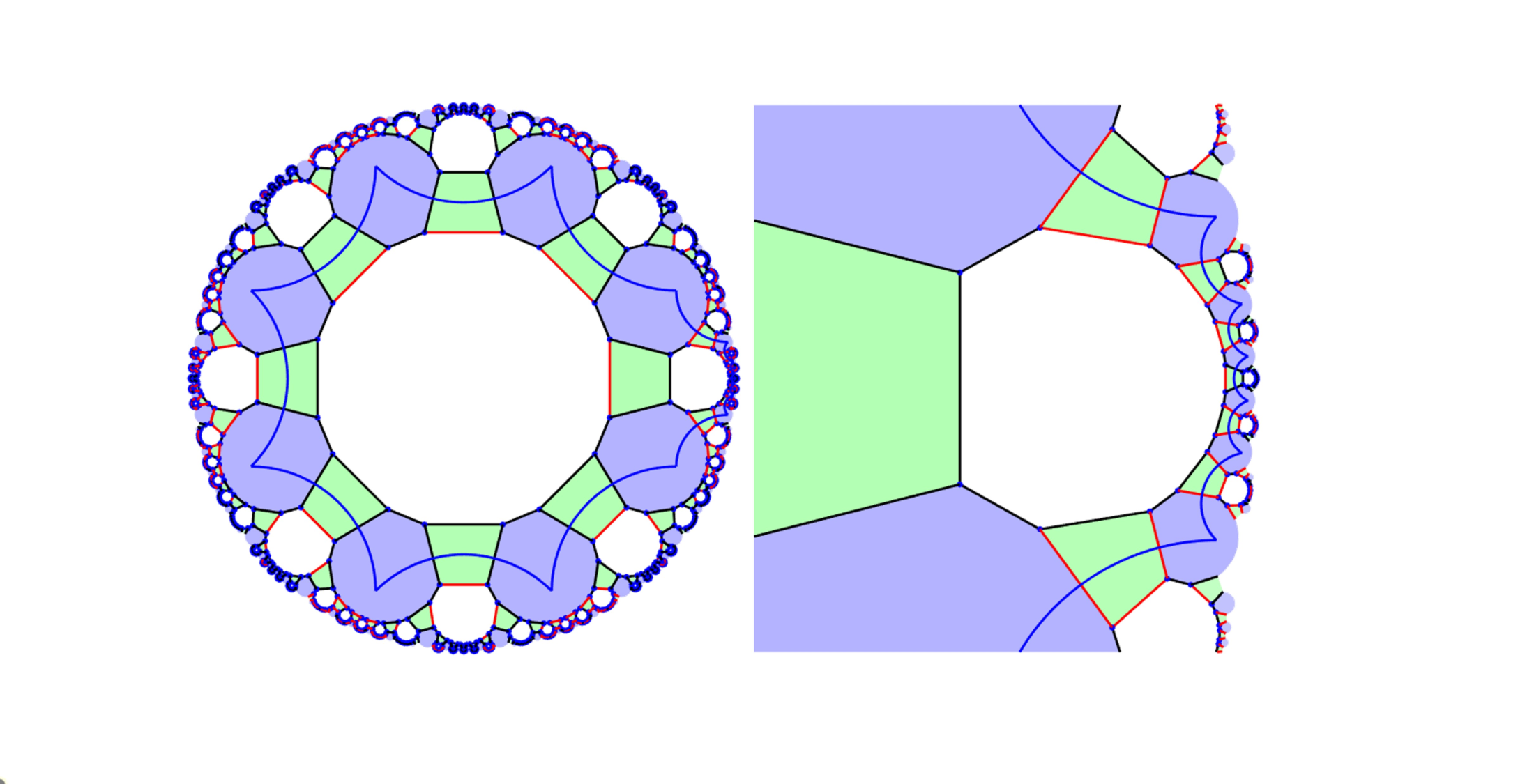}
	
	\caption{Top-down projection schematic of the tight-binding model with flux distribution $(\alpha_1, \alpha_2, \alpha_3, \beta_1, \beta_2, \beta_3)=(-1, -1, 1, 1, 1, 1)$. The black/red lines represent hopping coefficients $t_{ij}=1/-1$ for both top and bottom layers. The blue dots represent the hopping coefficients $t_{ij}=1$ between the top and bottom layers. The supercell is shown by blue curved lines.}
	\label{fig:phi12} 
\end{figure*}	
\subsection{tight-binding models with different fluxes}

For a given flux distribution, we can have different parameters of tight-binding models. For the $\pi$-flux shown in Fig. \ref{fig:flux_rotation}, as long as there exists odd number of $-1$ hopping coefficients on the surrounding edges, the flux condition will be satisfied. For the $\mathbb{Z}_2$ extended hyperbolic group $\tilde{\Delta}(2,8,8)$, there are 64 classes of symmetries which is predicted by the second cohomology group as shown in Eq.~(\ref{eq:H2ofhyperbolic}). These classes are labeled by $(\alpha_1, \alpha_2, \alpha_3, \beta_1, \beta_2, \beta_3)$ in Eq.~(\ref{eq:288relation}). We will list the 6 classes with only one parameter equals to -1 and the rest classes can be constructed by these 6 classes.

When $(\alpha_1, \alpha_2, \alpha_3, \beta_1, \beta_2, \beta_3)=(-1, 1, 1, 1, 1, 1)$, we show one example of the corresponding tight-binding model in Fig. \ref{fig:phi1}. The primitive cell has four pairs of periodic boundaries indicated by $\gamma_i, i=1, 2, 3, 4$, which is the same as the triangle group $\Delta(2,8,8)$ \cite{maciejko2022automorphic}. The four translation generators $\gamma_i$ form a subgroup $\Gamma$ of $\tilde{\Delta}(2,8,8)$ which is called the Fuchsian group:
		\begin{equation}
	\Gamma=\langle \gamma_1,\gamma_2,\gamma_3,\gamma_4|\gamma_1\gamma_2^{-1}\gamma_3\gamma_4^{-1}\gamma_1^{-1}\gamma_2\gamma_3^{-1}\gamma_4\rangle.
	\label{eq:fuchsian}
\end{equation}
Here the generators $\gamma_i$ are related to the generators of Eq.~(\ref{eq:triangle}) by $\gamma_i=(ca)^{5-i}ab(ca)^{i-1}$, which can be verified by the LINS package of the GAP \cite{LINS0.9}. The Fuchsian group $\Gamma$ is isomorphic to the fundamental group of a genus-2 surface, while the translation subgroups of wallpaper groups are isomorphic to the genus-1 surface.  

When $(\alpha_1, \alpha_2, \alpha_3, \beta_1, \beta_2, \beta_3)=(1, -1,1, 1, 1, 1)$, the tight-binding model is shown in Fig.~\ref{fig:phi2} where the supercell is twice the size of the primitive cell as shown in Fig.~\ref{fig:phi1}. According to the Riemann-Hurwitz formula \cite{lenggenhager2023non}, the relation between the genus of a supercell $g_{sc}$ and the genus of a primitive cell $g_{pc}$ can be expressed as:
\begin{equation}
	g_{sc}-1=N(g_{pc}-1),
	\label{eq:riemann}
\end{equation}
where $N$ is the number of primitive cells contained in the supercell. The seven pairs of periodic boundaries denoted by $\tilde{\gamma}_i$ are related to the translation operators $\gamma_i$ in Eq.~(\ref{eq:fuchsian}) by \cite{LINS0.9}:
\begin{subequations}
	\begin{eqnarray}
		\tilde{\gamma}_1=\gamma_1^2,
		\label{eq:phi2_1}
	\end{eqnarray}
	\begin{eqnarray}
		\tilde{\gamma}_2=\gamma_2,
		\label{eq:phi2_2}
	\end{eqnarray}
	\begin{eqnarray}
		\tilde{\gamma}_3=\gamma_3,
		\label{eq:phi2_3}
	\end{eqnarray}
	\begin{eqnarray}
		\tilde{\gamma}_4=\gamma_4,
		\label{eq:phi2_4}
	\end{eqnarray}
	\begin{eqnarray}
		\tilde{\gamma}_5=\gamma_1^{-1}\gamma_2\gamma_1,
		\label{eq:phi2_5}
	\end{eqnarray}
	\begin{eqnarray}
		\tilde{\gamma}_6=\gamma_1^{-1}\gamma_3\gamma_1,
		\label{eq:phi2_6}
	\end{eqnarray}
		\begin{eqnarray}
		\tilde{\gamma}_{p_1}=\gamma_1^{-1}\gamma_4\gamma_1=\tilde{\gamma}_6\tilde{\gamma}_5^{-1}\tilde{\gamma}_4\tilde{\gamma}_3^{-1}\tilde{\gamma}_2.
		\label{eq:phi2_7}
	\end{eqnarray}
	\label{eq:phi2}
\end{subequations}
Six translation generators $\tilde{\gamma}_i$ shown in Eq.~(\ref{eq:phi2}) are independent of each other, which satisfies the Riemann-Hurwitz formula shown in Eq.~(\ref{eq:riemann}) where $N=2, g_{pc}=2$. The translation group $\Gamma^{(2)}_A$ generated by the $\tilde{\gamma}_i$ is isomorphic to the fundamental group of a genus-3 surface.
	
When $(\alpha_1, \alpha_2, \alpha_3, \beta_1, \beta_2, \beta_3)=(1, 1,-1, 1, 1, 1)$, the tight-binding model is shown in Fig. \ref{fig:phi3} (see Appendix~\ref{appendix:tight_binding}). The periodic boundary conditions are exactly the same as the model in Fig. \ref{fig:phi2}.

When $(\alpha_1, \alpha_2, \alpha_3, \beta_1, \beta_2, \beta_3)=(1, 1, 1, -1, 1, 1)$, the tight-binding model is shown in Fig. \ref{fig:phi4}, where the size of the supercell is also twice the size of the primitive cell shown in Fig. \ref{fig:phi1}. The eight pairs of periodic boundaries $\tilde{\gamma}_i$ are related to $\gamma_i$ in Eq.~(\ref{eq:fuchsian}) by:
\begin{subequations}
	\begin{eqnarray}
		\tilde{\gamma}_1=\gamma_4\gamma_1,
		\label{eq:phi4_1}
	\end{eqnarray}
	\begin{eqnarray}
		\tilde{\gamma}_2=\gamma_1^{-1}\gamma_2,
		\label{eq:phi4_2}
	\end{eqnarray}
	\begin{eqnarray}
		\tilde{\gamma}_3=\gamma_2^{-1}\gamma_3,
		\label{eq:phi4_3}
	\end{eqnarray}
	\begin{eqnarray}
		\tilde{\gamma}_4=\gamma_3^{-1}\gamma_4,
		\label{eq:phi4_4}
	\end{eqnarray}
	\begin{eqnarray}
		\tilde{\gamma}_5=\gamma_4^{-1}\gamma_1^{-1},
		\label{eq:phi4_5}
	\end{eqnarray}
	\begin{eqnarray}
		\tilde{\gamma}_6=\gamma_1\gamma_2^{-1},
		\label{eq:phi4_6}
	\end{eqnarray}
	\begin{eqnarray}
		\tilde{\gamma}_{p_1}=\gamma_2\gamma_3^{-1}=\tilde{\gamma}_5^{-1}\tilde{\gamma}_3^{-1}\tilde{\gamma}_1^{-1},
		\label{eq:phi4_7}
	\end{eqnarray}
		\begin{eqnarray}
		\tilde{\gamma}_{p_2}=\gamma_3\gamma_4^{-1}=\tilde{\gamma}_6^{-1}\tilde{\gamma}_4^{-1}\tilde{\gamma}_2^{-1},
		\label{eq:phi4_8}
	\end{eqnarray}
	\label{eq:phi4}
\end{subequations}
Again, the six independent translation generators $\tilde{\gamma}_i$ in Eq.~(\ref{eq:phi4}) indicate that the translation group $\Gamma^{(2)}_B$ they formed is isomorphic to the fundamental group of a genus-3 surface, which verifies the Riemann-Hurwitz formula.
For the cases $(\alpha_1, \alpha_2, \alpha_3, \beta_1, \beta_2, \beta_3)=(1, 1, 1, 1, -1, 1)$ and $(\alpha_1, \alpha_2, \alpha_3, \beta_1, \beta_2, \beta_3)=(1, 1, 1, 1, 1, -1)$, we list them in Appendix~\ref{appendix:tight_binding} due to the similarity with Fig. \ref{fig:phi4}.
	
After the 6 classes with only one parameter equals to -1 are equipped, we can combine any number of them to create the whole 64 classes of symmetries predicted in Eq.~(\ref{eq:H2ofhyperbolic}). For example, considering the case when $(\alpha_1, \alpha_2, \alpha_3, \beta_1, \beta_2, \beta_3)=(-1, -1, 1, 1, 1, 1)$, we can view it as the combination of the class $(\alpha_1, \alpha_2, \alpha_3, \beta_1, \beta_2, \beta_3)=(-1, 1, 1, 1, 1, 1)$ and $(\alpha_1, \alpha_2, \alpha_3, \beta_1, \beta_2, \beta_3)=(1, -1, 1, 1, 1, 1)$. The translation group  of the class $(\alpha_1, \alpha_2, \alpha_3, \beta_1, \beta_2, \beta_3)=(-1, -1, 1, 1, 1, 1)$ should be the subgroup of both $\Gamma$ shown in Fig. \ref{fig:phi1} and $\Gamma^{(2)}_A$ shown in Fig. \ref{fig:phi2}. Since $\Gamma \triangleright \Gamma^{(2)}_A$, the supercell of $(\alpha_1, \alpha_2, \alpha_3, \beta_1, \beta_2, \beta_3)=(-1, -1, 1, 1, 1, 1)$ shares the same shape with the supercell of group $\Gamma^{(2)}_A$, which is shown in Fig. \ref{fig:phi12}. The hopping coefficients $t_{ij}^{(-1,-1,1,1,1,1)}$ inside the supercell are the products of the corresponding generators from the six listed classes:
$t_{ij}^{(-1,-1,1,1,1,1)}=t_{ij}^{(-1,1,1,1,1,1)}t_{ij}^{(1,-1,1,1,1,1)}$. We can verify that the hopping coefficients in Fig. \ref{fig:phi12} are created by the products of the hopping coefficients in Fig. \ref{fig:phi1} and Fig. \ref{fig:phi2}.
	
When $(\alpha_1, \alpha_2, \alpha_3, \beta_1, \beta_2, \beta_3)=(1, -1, 1, -1, 1, 1)$, the translation group $\Gamma^{(3)}$ of the supercell as shown in Fig. \ref{fig:phi24} (see Appendix~\ref{appendix:tight_binding}) should be the normal subgroup of both $\Gamma^{(2)}_A$ and $\Gamma^{(2)}_B$ described by Eq.~(\ref{eq:phi2}) and Eq.~(\ref{eq:phi4}) respectively. The 17 pairs of periodic boundaries denoted by $\tilde{\gamma_i}$ are related to the $\gamma_i$ in Eq.~(\ref{eq:fuchsian}) by:
\begin{subequations}
	\begin{eqnarray}
		\tilde{\gamma}_1=\gamma_1^{-2},
		\label{eq:phi24_1}
	\end{eqnarray}
	\begin{eqnarray}
		\tilde{\gamma}_2=\gamma_3^{-1}\gamma_4,
		\label{eq:phi24_2}
	\end{eqnarray}
	\begin{eqnarray}
		\tilde{\gamma}_3=\gamma_2\gamma_4,
		\label{eq:phi24_3}
	\end{eqnarray}
	\begin{eqnarray}
		\tilde{\gamma}_4=\gamma_2^{-1}\gamma_3,
		\label{eq:phi24_4}
	\end{eqnarray}
	\begin{eqnarray}
		\tilde{\gamma}_5=\gamma_1^{-1}\gamma_2^{-1}\gamma_3\gamma_1,
		\label{eq:phi24_5}
	\end{eqnarray}
	\begin{eqnarray}
		\tilde{\gamma}_6=\gamma_1^{-1}\gamma_4\gamma_2\gamma_1,
		\label{eq:phi24_6}
	\end{eqnarray}
		\begin{eqnarray}
		\tilde{\gamma}_7=\gamma_1^{-1}\gamma_3\gamma_4^{-1}\gamma_1,
		\label{eq:phi24_7}
	\end{eqnarray}
	\begin{eqnarray}
		\tilde{\gamma}_8=\gamma_1^{-1}\gamma_2\gamma_3^{-1}\gamma_1,
		\label{eq:phi24_8}
	\end{eqnarray}
		\begin{eqnarray}
		\tilde{\gamma}_9=\gamma_3\gamma_4^{-1},
		\label{eq:phi24_9}
	\end{eqnarray}
			\begin{eqnarray}
		\tilde{\gamma}_{10}=\gamma_2\gamma_3^{-1},
		\label{eq:phi24_10}
	\end{eqnarray}
	\begin{eqnarray}
		\tilde{\gamma}_{p_1}=\tilde{\gamma}_6^{-1}\tilde{\gamma}_8^{-1}\tilde{\gamma}_9^{-1}\tilde{\gamma}_3,
		\label{eq:phi24_p1}
	\end{eqnarray}
		\begin{eqnarray}
		\tilde{\gamma}_{p_2}=\tilde{\gamma}_5^{-1}\tilde{\gamma}_1\tilde{\gamma}_2^{-1},
		\label{eq:phi24_p2}
	\end{eqnarray}
	\begin{eqnarray}
		\tilde{\gamma}_{p_3}=\tilde{\gamma}_8^{-1}\tilde{\gamma}_9^{-1},
		\label{eq:phi24_p3}
	\end{eqnarray}
		\begin{eqnarray}
		\tilde{\gamma}_{p_4}=\tilde{\gamma}_7\tilde{\gamma}_1\tilde{\gamma}_{10},
		\label{eq:phi24_p4}
	\end{eqnarray}
	\begin{eqnarray}
		\tilde{\gamma}_{p_5}=\tilde{\gamma}_9^{-1}\tilde{\gamma}_{10}^{-1}\tilde{\gamma}_3\tilde{\gamma}_2^{-1}\tilde{\gamma}_4^{-1},
		\label{eq:phi24_p5}
	\end{eqnarray}
		\begin{eqnarray}
		\tilde{\gamma}_{p_6}=\tilde{\gamma}_6^{-1}\tilde{\gamma}_8^{-1}\tilde{\gamma}_9^{-1}\tilde{\gamma}_3\tilde{\gamma}_4^{-1},
		\label{eq:phi24_p6}
	\end{eqnarray}
	\begin{eqnarray}
		\tilde{\gamma}_{p_7}=\tilde{\gamma}_8\tilde{\gamma}_7\tilde{\gamma}_6\tilde{\gamma}_5\tilde{\gamma}_6^{-1}\tilde{\gamma}_8^{-1}\tilde{\gamma}_9^{-1}\tilde{\gamma}_3\tilde{\gamma}_4^{-1}.
		\label{eq:phi24_p7}
	\end{eqnarray}
	\label{eq:phi24}
\end{subequations}
The translation group $\Gamma^{(3)}$ formed by ten independent translation generators $\tilde{\gamma}_i$ in Eq.~(\ref{eq:phi24}) is isomorphic to the fundamental group of a genus-5 surface, which can be predicted by the Riemann-Hurwitz formula in Eq.~(\ref{eq:riemann}) since $N=4, g_{pc}=2$.
The relations between the Fuchsian group $\Gamma$ (Eq.~(\ref{eq:fuchsian})), the translation groups $\Gamma^{(2)}_A, \Gamma^{(2)}_B$ (Eq.~(\ref{eq:phi2}), Eq.~(\ref{eq:phi4})), and the translation group $\Gamma^{(3)}$ (Eq.~(\ref{eq:phi24})) can be concluded by the following sequences of finite-index normal subgroups \cite{lenggenhager2023non}:
\begin{subequations}
	\begin{eqnarray}
		\Gamma \triangleright \Gamma^{(2)}_A \triangleright \Gamma^{(3)},
	\end{eqnarray}
	\begin{eqnarray}
		\Gamma \triangleright \Gamma^{(2)}_B \triangleright \Gamma^{(3)}.
	\end{eqnarray}
\end{subequations}
The indices of the quotient groups are $|\Gamma/\Gamma^{(2)}_{A,B}|=2$, $|\Gamma^{(2)}_{A,B}/\Gamma^{(3)}|=2$, which can be verified by counting the number of primitive cells in the supercells shown in Fig. \ref{fig:phi2}, Fig. \ref{fig:phi4}, and Fig. \ref{fig:phi24}. 

\section{Band theory in $\mathbb{Z}_2$ extended hyperbolic lattices}
\begin{figure}
	\centering
	\includegraphics[width=.5\textwidth]{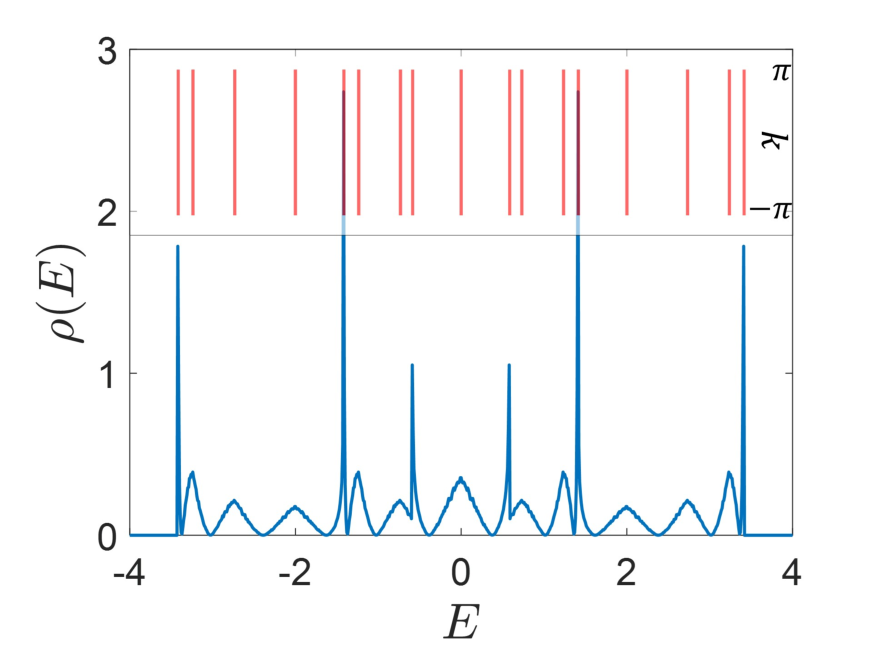}
	
	\caption{Density of states $\rho(E)$ of the tight-binding model on the hyperbolic lattice with flux distribution $(\alpha_1, \alpha_2, \alpha_3, \beta_1, \beta_2, \beta_3)=(-1, 1, 1, 1, 1, 1)$. The inset shows the all-flat dispersions when $k_1=k_2=k_3=k_4=k$ in Eq.~(\ref{eq:abelian_HBT}).}
	\label{fig:dos1} 
\end{figure}

For simplicity, we assume the eigenstates of the Hamiltonian satisfy the Abelian hyperbolic band theory \cite{maciejko2021hyperbolic}
	\begin{equation}
		\psi ( \gamma^{-1}_j z_i)=e^{ik_j}\psi(z_i),
		\label{eq:abelian_HBT}
	\end{equation}
where $z_i$ is the site on the lattice and $\gamma_j$ are the translation operators in the primitive cell. Similarly, we can replace the translation operators $\gamma_j$ with $\tilde{\gamma}_j$ for the Bloch theory in the supercells. Although the translation generators are non-Abelian ($\gamma_i\gamma_j\neq\gamma_j\gamma_i$) for the Fuchsian group $\Gamma$, the Abelian hyperbolic band theory can be applied to describe the abelianization of $\Gamma$, which is $\Gamma/[\Gamma,\Gamma]$ \cite{lux2023converging}. The theory is widely adopted and proven effective in prior studies \cite{boettcher2022crystallography,chen2023hyperbolic,chen2024anomalous,yuan2024hyperbolic}. By applying Eq.~(\ref{eq:abelian_HBT}) to the tight-binding model shown in Fig. \ref{fig:phi1}, we can achieve the density of states (DOS) $\rho(E)$ of the hyperbolic lattice with flux distribution $(\alpha_1, \alpha_2, \alpha_3, \beta_1, \beta_2, \beta_3)=(-1, 1, 1, 1, 1, 1)$ as shown in Fig. \ref{fig:dos1}. Also, the all-flat dispersion relations are shown when $k_1=k_2=k_3=k_4=k$ in Eq.~(\ref{eq:abelian_HBT}). Note that there exist 32 sites in the primitive cell of Fig. \ref{fig:phi1}, which means some bands shown in Fig. \ref{fig:dos1} are degenerate since only 17 isolated lines are presented. The positions of the van Hove singularities in DOS match well with the discrete eigenenergies of the flat bands. Similar all-flat bands along a single specified line in momentum space are also reported in recent references \cite{chan2024superconductivity,ara2025flat,fukui2025topological}. The DOS of tight-binding models with other flux distributions are shown in Appendix~\ref{appendix:dos}.

\section{Discussion and outlook}
In summary, this work demonstrates that $\mathbb{Z}_2$ gauge extensions of hyperbolic lattices can be systematically classified and realized through nearest-neighbor tight-binding models. The resulting lattice models reveal a hierarchy of translation normal subgroups which are associated with high genus surfaces according to the Riemann-Hurwitz formula. The emergence of all-flat dispersions along specific directions in momentum space highlights the unique property of the hyperbolic lattice with the $\mathbb{Z}_2$ gauge structure. Our classification framework and model construction can be extended to other non-Euclidean lattices and gauge structures such as $\mathbb{Z}_n$ or $U(1)$ groups. Overall, our results provide an approach to understanding how gauge fields and hyperbolic geometry intertwine to yield new classes of lattice models, paving the way for future studies of geometry-induced fractionalization and topological order beyond Euclidean space.

	\appendix
\section{\label{appendix:cohomology}Classify projective representations of a group by the second cohomology group}

Here, we want to classify the projective representations of the group $G$ with coefficients in the Abelian group $A$ by the second cohomology group $H^2(G,A)$. For clarity, we reproduce here Eq.~(\ref{eq:modified_multiplication}) from the main text:
\begin{equation}
	\tilde{\rho}(g_1)\tilde{\rho}(g_2)=\omega(g_1,g_2)\tilde{\rho}(g_1g_2),\quad  \forall g_1,g_2\in G.
	\label{eq:modified_multiplication2}
\end{equation}
The scalars $\omega(g_1,g_2)\in A$ satisfy the 2-cocycle condition due to the associativity of group multiplication:
\begin{equation}
	\omega(g_1,g_2)\omega(g_1g_2,g_3)=\omega(g_2,g_3)\omega(g_1,g_2g_3).
\end{equation}
All possible 2-cocycles form a group $Z^2(G,A)$ under pointwise multiplication.
A gauge transformation can be applied to $\tilde{\rho}$ and the transformed projective representation can be expressed as:
\begin{equation}
	\tilde{\rho}\prime(g)=\chi(g)\tilde{\rho}(g),
	\label{eq:gauged_rho}
\end{equation}
where gauge $\chi(g)\in A$. Combining Eq. (\ref{eq:modified_multiplication2}) and Eq. (\ref{eq:gauged_rho}) we can obtain:
\begin{equation}
	\omega\prime(g_1,g_2)=\omega(g_1,g_2)\frac{\chi(g_1)\chi(g_2)}{\chi(g_1g_2)}.
\end{equation}
Here $\omega\prime$ and $\omega$ are considered to be equivalent if they are related by a 2-cocycle $\frac{\chi(g_1)\chi(g_2)}{\chi(g_1g_2)}$. All the trivial 2-cocycles generated by the gauges $\chi(g)$ form an abelian group $B^2(G,A)\subset Z^2(G,A)$. The second cohomology group $H^2(G,A)$ can be defined as the quotient group of $B^2(G,A)$ in $Z^2(G,A)$:
\begin{equation}
	H^2(G,A)=Z^2(G,A)/B^2(G,A).
	\label{eq:H2_def}
\end{equation}

\section{\label{appendix:tight_binding}tight-binding model with different fluxes}
\begin{figure*}
	\centering
	
	\includegraphics[width=\textwidth]{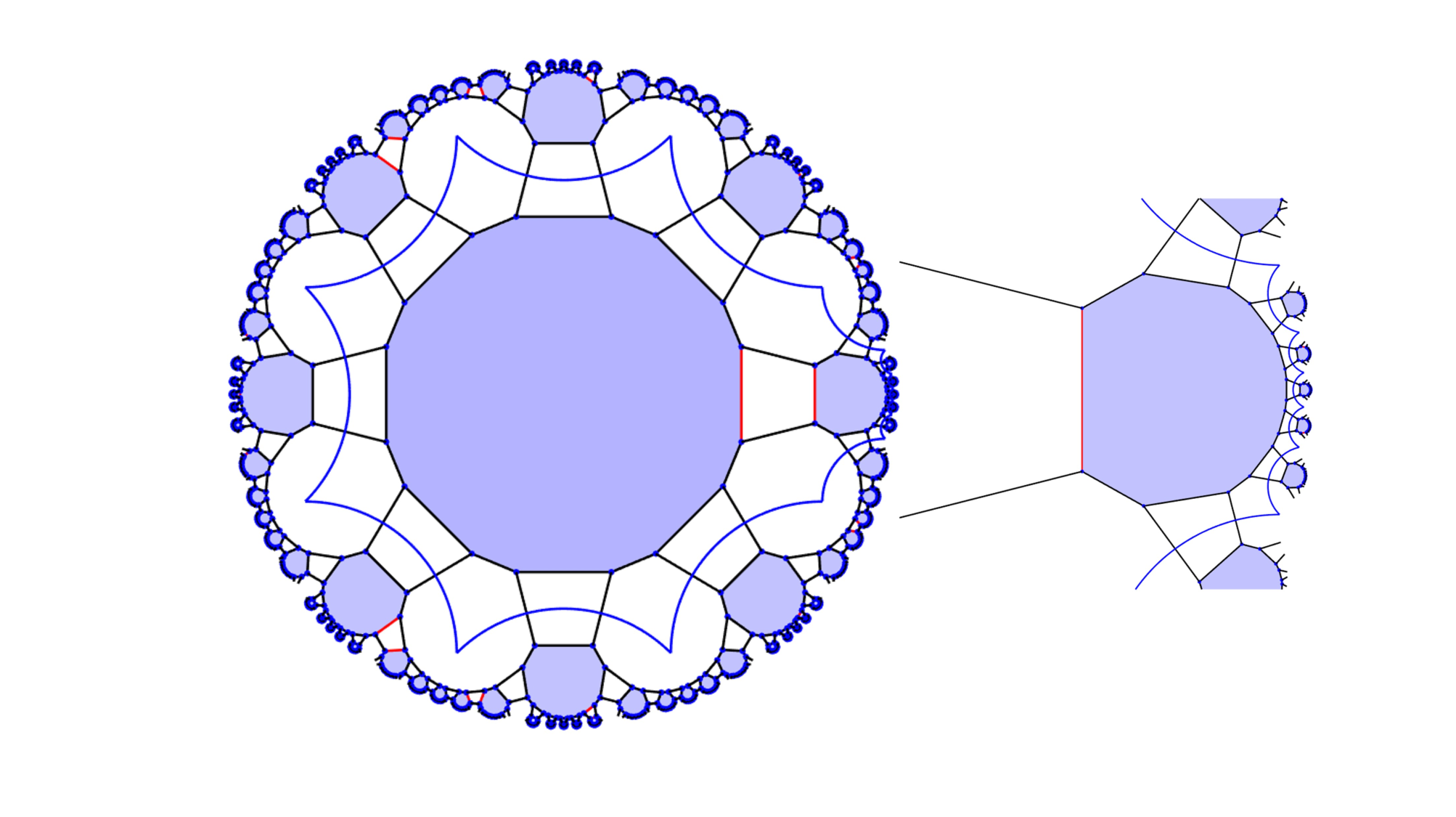}
	
	\caption{Top-down projection schematic of the tight-binding model with flux distribution $(\alpha_1, \alpha_2, \alpha_3, \beta_1, \beta_2, \beta_3)=(1, 1, -1, 1, 1, 1)$. The black/red lines represent hopping coefficients $t_{ij}=1/-1$ for both top and bottom layers. The blue dots represent the hopping coefficients $t_{ij}=1$ between the top and bottom layers. The supercell is shown by blue curved lines.}
	\label{fig:phi3} 
\end{figure*}

When $(\alpha_1, \alpha_2, \alpha_3, \beta_1, \beta_2, \beta_3)=(1, 1,-1, 1, 1, 1)$, the tight-binding model is shown in Fig. \ref{fig:phi3}.

\begin{figure}
		\centering
		\includegraphics[width=0.55\textwidth]{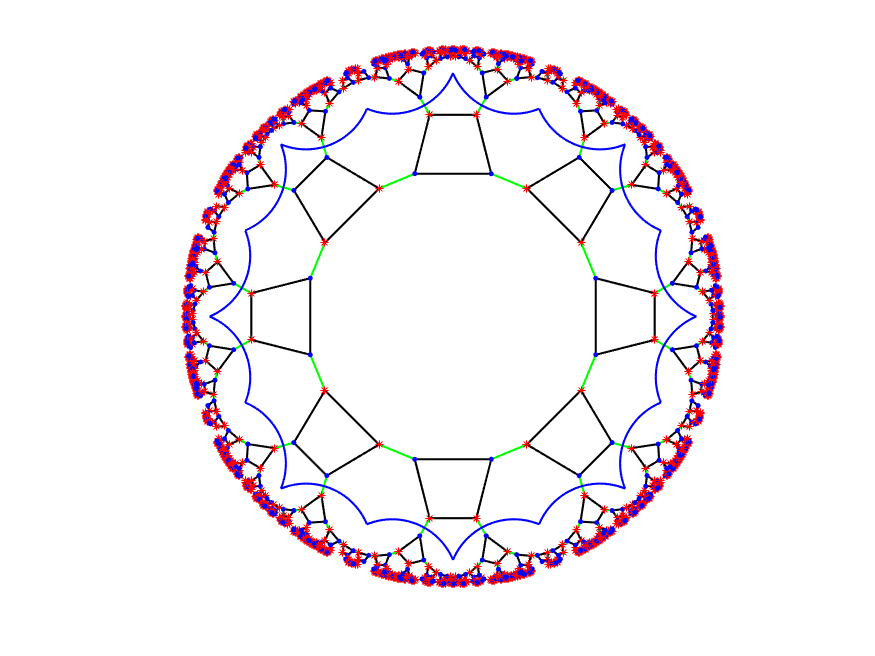}
		\caption{Top-down projection schematic of the tight-binding model with flux distribution $(\alpha_1, \alpha_2, \alpha_3, \beta_1, \beta_2, \beta_3)=(1, 1, 1, 1, -1, 1)$. The black lines represent hopping coefficients $t_{ij}=1$ for both top and bottom layers while the green lines represent hopping coefficients $t_{ij}=1$ for the top layer and $t_{ij}=-1$ for the bottom layer. The blue dots/red stars represent the hopping coefficients $t_{ij}=1/-1$ between the top and bottom layers. The supercell is shown by blue curved lines.}
		\label{fig:phi5} 
	\end{figure}

When $(\alpha_1, \alpha_2, \alpha_3, \beta_1, \beta_2, \beta_3)=(1, 1,1, 1, -1, 1)$, the tight-binding model is shown in Fig. \ref{fig:phi5}.
	
\begin{figure}
	\centering
	\includegraphics[width=0.55\textwidth]{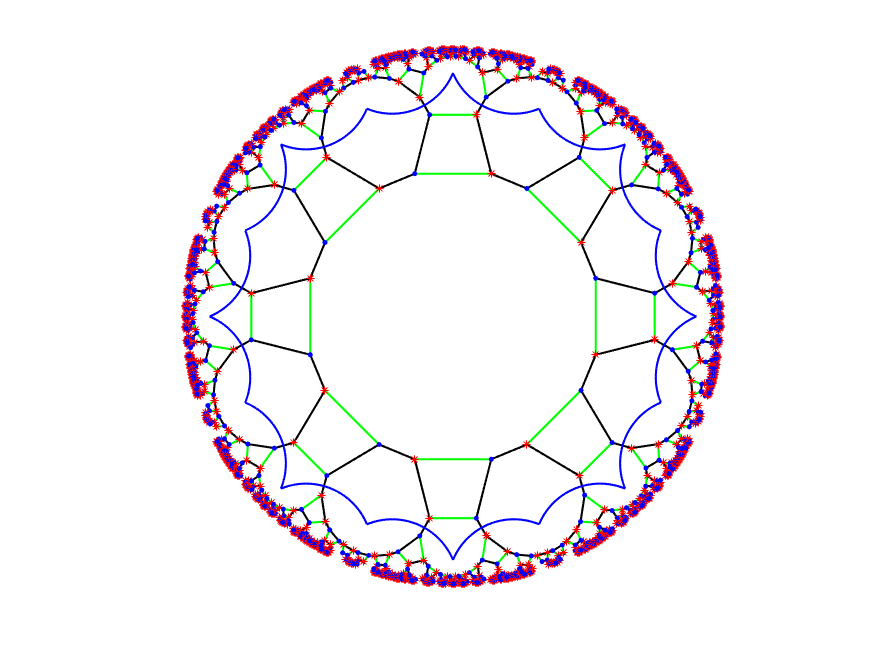}
	\caption{Top-down projection schematic of the tight-binding model with flux distribution $(\alpha_1, \alpha_2, \alpha_3, \beta_1, \beta_2, \beta_3)=(1, 1, 1, 1, 1, -1)$. The black lines represent hopping coefficients $t_{ij}=1$ for both top and bottom layers while the green lines represent hopping coefficients $t_{ij}=1$ for the top layer and $t_{ij}=-1$ for the bottom layer. The blue dots/red stars represent the hopping coefficients $t_{ij}=1/-1$ between the top and bottom layers. The supercell is shown by blue curved lines.}
	\label{fig:phi6} 
\end{figure}

When $(\alpha_1, \alpha_2, \alpha_3, \beta_1, \beta_2, \beta_3)=(1, 1,1, 1, 1, -1)$, the tight-binding model is shown in Fig. \ref{fig:phi6}.

\begin{figure*}
	\centering
	
	\includegraphics[width=1.3\textwidth]{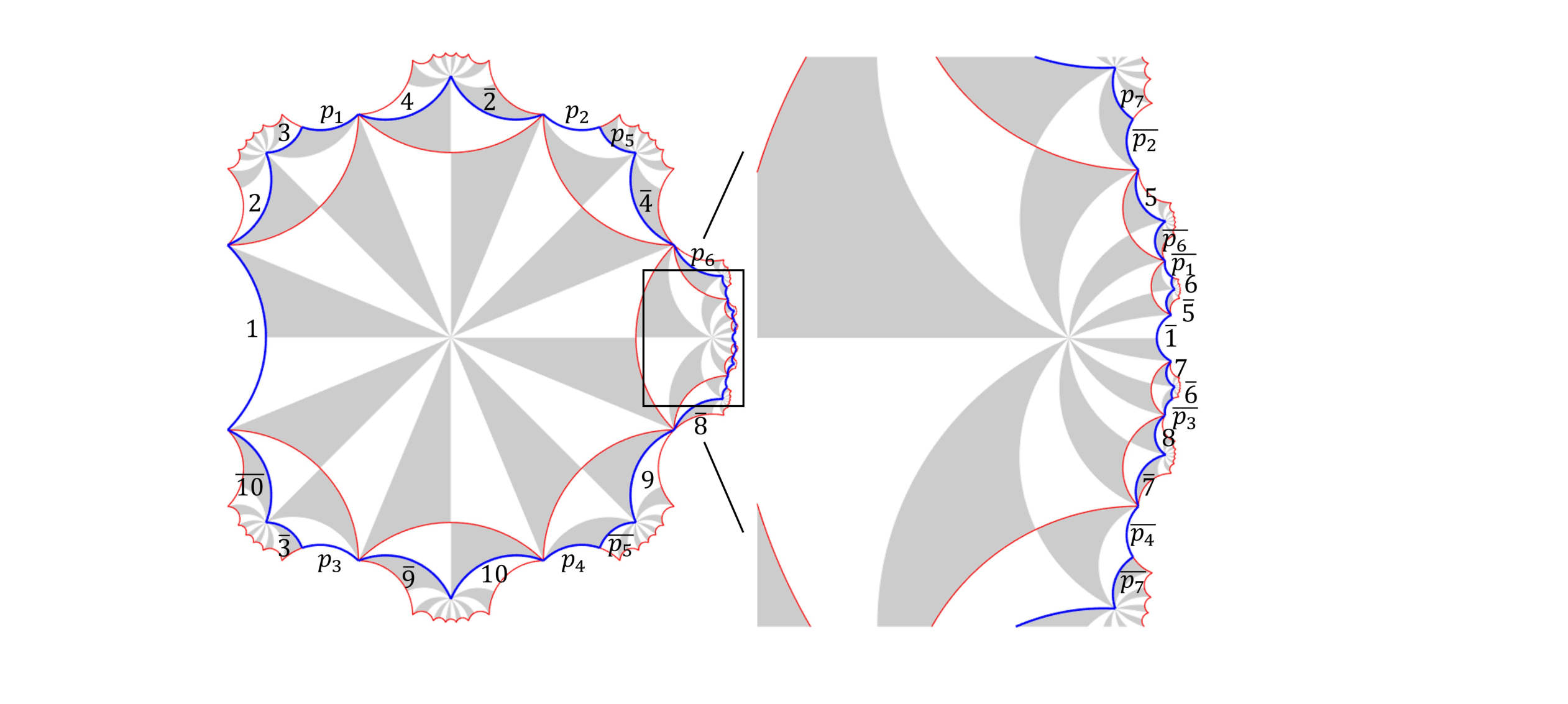}
	
	\caption{The blue curved lines represent the supercell when $(\alpha_1, \alpha_2, \alpha_3, \beta_1, \beta_2, \beta_3)=(1, -1, 1, -1, 1, 1)$ while the red curved lines represent the primitive cell. 17 pairs of periodic boundaries $\bar{i}$ and $i$ are related by the translation operators $\tilde{\gamma_i}$, where 10 of them are independent.}
	\label{fig:phi24} 
\end{figure*}

When $(\alpha_1, \alpha_2, \alpha_3, \beta_1, \beta_2, \beta_3)=(1, -1,1, -1, 1, 1)$, the tight-binding model is shown in Fig. \ref{fig:phi24}.

\section{\label{appendix:dos}Density of states of tight-binding models with different flux distributions }
\begin{figure}
	\centering
	
	\subfloat[]{{\includegraphics[width=.25\textwidth]{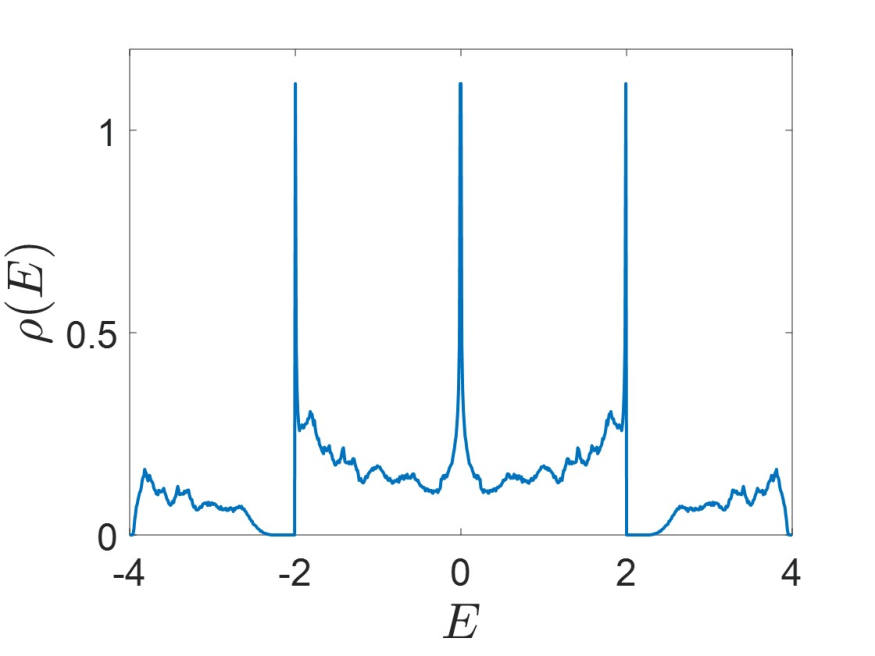}}\label{fig:dos2}}
	\subfloat[]{{\includegraphics[width=.25\textwidth]{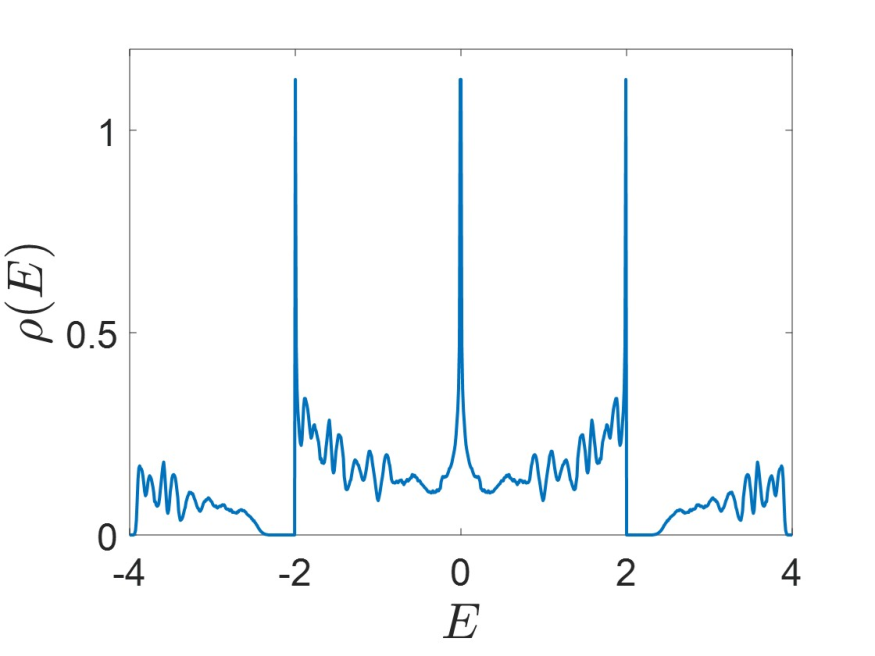}}\label{fig:dos3}}\\
	\subfloat[]{{\includegraphics[width=.25\textwidth]{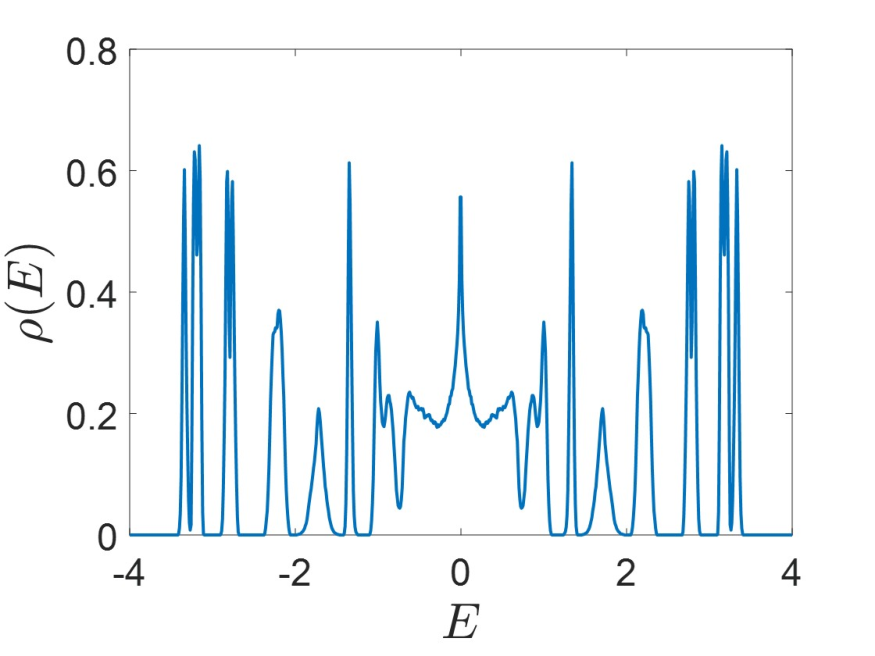}}\label{fig:dos4}}
	\subfloat[]{{\includegraphics[width=.25\textwidth]{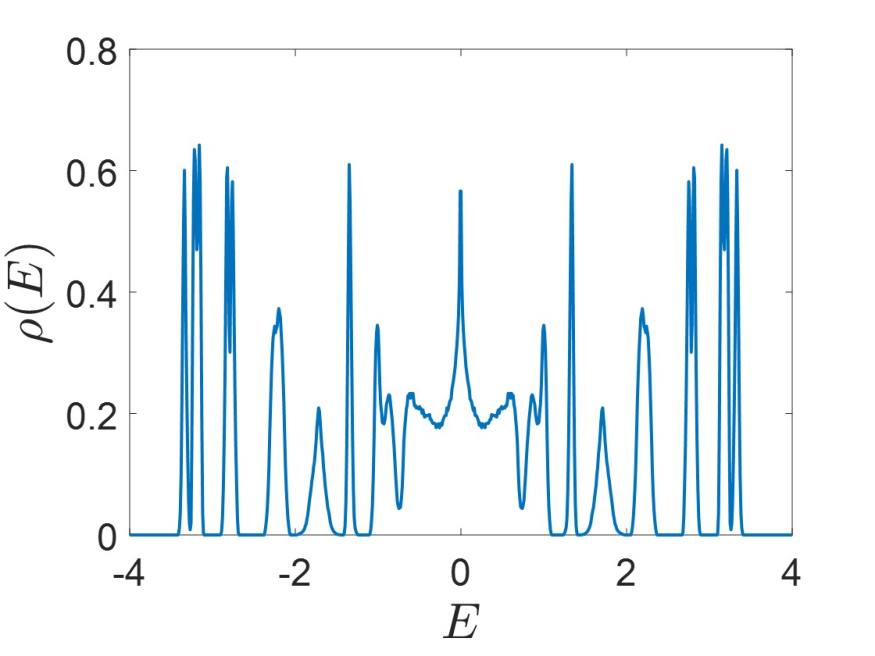}}\label{fig:dos5}}\\
	\subfloat[]{{\includegraphics[width=.25\textwidth]{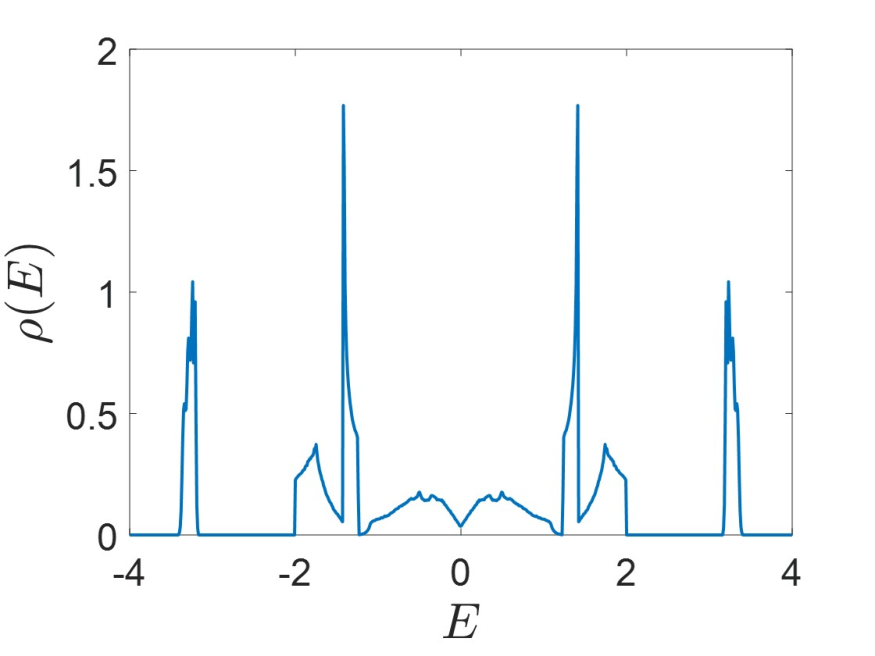}}\label{fig:dos6}}
	\subfloat[]{{\includegraphics[width=.25\textwidth]{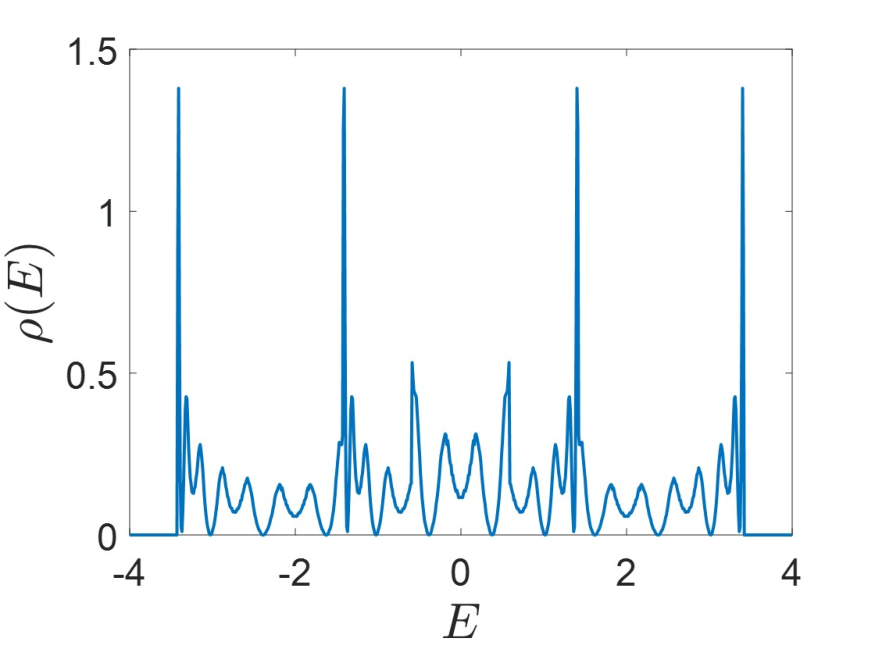}}\label{fig:dos12}}\\
	\caption{Density of states $\rho(E)$ of the tight-binding models on the hyperbolic lattices with flux distributions (a) $(\alpha_1, \alpha_2, \alpha_3, \beta_1, \beta_2, \beta_3)=(1, -1, 1, 1, 1, 1)$, (b) $(\alpha_1, \alpha_2, \alpha_3, \beta_1, \beta_2, \beta_3)=(1, 1, -1, 1, 1, 1)$, (c) $(\alpha_1, \alpha_2, \alpha_3, \beta_1, \beta_2, \beta_3)=(1, 1, 1, -1, 1, 1)$, (d) $(\alpha_1, \alpha_2, \alpha_3, \beta_1, \beta_2, \beta_3)=(1, 1, 1, 1, -1, 1)$, (e) $(\alpha_1, \alpha_2, \alpha_3, \beta_1, \beta_2, \beta_3)=(1, 1, 1, 1, 1, -1)$, (f) $(\alpha_1, \alpha_2, \alpha_3, \beta_1, \beta_2, \beta_3)=(-1, -1, 1, 1, 1, 1)$. }
	\label{fig:dos_appendix} 
\end{figure}

The DOS of tight-binding models with different flux distributions are shown in Fig. \ref{fig:dos_appendix} , where 5 of them are from the 6 classes with only one parameter equals to -1 and the rest is the combination of them corresponding to Fig. \ref{fig:phi12}. Due to the symmetry between generators $\tilde{a}$ and $\tilde{b}$ shown in the relations (see Eq.~(\ref{eq:288relation})), the similarity between the DOS shown in Fig. \ref{fig:dos4} and Fig. \ref{fig:dos5} is expected.

	\begin{acknowledgments}
C.W.Q. acknowledges the support from the Ministry of Education, Singapore (Grant No. A-8002152-00-00)  and the
Science and Technology Project of Jiangsu Province (Grant
No. BZ2022056). The authors would like to acknowledge
the support from the Research Platform for Energy and
Environmental Nanotech, National University of Singapore
(Suzhou) Research Institute.
	\end{acknowledgments}
	
\nocite{*}
\bibliography{hyperbolic}

\providecommand{\noopsort}[1]{}\providecommand{\singleletter}[1]{#1}%
\begin{thebibliography}{27}%
\makeatletter
\providecommand \@ifxundefined [1]{%
 \@ifx{#1\undefined}
}%
\providecommand \@ifnum [1]{%
 \ifnum #1\expandafter \@firstoftwo
 \else \expandafter \@secondoftwo
 \fi
}%
\providecommand \@ifx [1]{%
 \ifx #1\expandafter \@firstoftwo
 \else \expandafter \@secondoftwo
 \fi
}%
\providecommand \natexlab [1]{#1}%
\providecommand \enquote  [1]{``#1''}%
\providecommand \bibnamefont  [1]{#1}%
\providecommand \bibfnamefont [1]{#1}%
\providecommand \citenamefont [1]{#1}%
\providecommand \href@noop [0]{\@secondoftwo}%
\providecommand \href [0]{\begingroup \@sanitize@url \@href}%
\providecommand \@href[1]{\@@startlink{#1}\@@href}%
\providecommand \@@href[1]{\endgroup#1\@@endlink}%
\providecommand \@sanitize@url [0]{\catcode `\\12\catcode `\$12\catcode
  `\&12\catcode `\#12\catcode `\^12\catcode `\_12\catcode `\%12\relax}%
\providecommand \@@startlink[1]{}%
\providecommand \@@endlink[0]{}%
\providecommand \url  [0]{\begingroup\@sanitize@url \@url }%
\providecommand \@url [1]{\endgroup\@href {#1}{\urlprefix }}%
\providecommand \urlprefix  [0]{URL }%
\providecommand \Eprint [0]{\href }%
\providecommand \doibase [0]{https://doi.org/}%
\providecommand \selectlanguage [0]{\@gobble}%
\providecommand \bibinfo  [0]{\@secondoftwo}%
\providecommand \bibfield  [0]{\@secondoftwo}%
\providecommand \translation [1]{[#1]}%
\providecommand \BibitemOpen [0]{}%
\providecommand \bibitemStop [0]{}%
\providecommand \bibitemNoStop [0]{.\EOS\space}%
\providecommand \EOS [0]{\spacefactor3000\relax}%
\providecommand \BibitemShut  [1]{\csname bibitem#1\endcsname}%
\let\auto@bib@innerbib\@empty
\bibitem [{\citenamefont {Maciejko}\ and\ \citenamefont
  {Rayan}(2021)}]{maciejko2021hyperbolic}%
  \BibitemOpen
  \bibfield  {author} {\bibinfo {author} {\bibfnamefont {J.}~\bibnamefont
  {Maciejko}}\ and\ \bibinfo {author} {\bibfnamefont {S.}~\bibnamefont
  {Rayan}},\ }\href@noop {} {\bibfield  {journal} {\bibinfo  {journal} {Science
  advances}\ }\textbf {\bibinfo {volume} {7}},\ \bibinfo {pages} {eabe9170}
  (\bibinfo {year} {2021})}\BibitemShut {NoStop}%
\bibitem [{\citenamefont {Boettcher}\ \emph {et~al.}(2022)\citenamefont
  {Boettcher}, \citenamefont {Gorshkov}, \citenamefont {Koll{\'a}r},
  \citenamefont {Maciejko}, \citenamefont {Rayan},\ and\ \citenamefont
  {Thomale}}]{boettcher2022crystallography}%
  \BibitemOpen
  \bibfield  {author} {\bibinfo {author} {\bibfnamefont {I.}~\bibnamefont
  {Boettcher}}, \bibinfo {author} {\bibfnamefont {A.~V.}\ \bibnamefont
  {Gorshkov}}, \bibinfo {author} {\bibfnamefont {A.~J.}\ \bibnamefont
  {Koll{\'a}r}}, \bibinfo {author} {\bibfnamefont {J.}~\bibnamefont
  {Maciejko}}, \bibinfo {author} {\bibfnamefont {S.}~\bibnamefont {Rayan}},\
  and\ \bibinfo {author} {\bibfnamefont {R.}~\bibnamefont {Thomale}},\
  }\href@noop {} {\bibfield  {journal} {\bibinfo  {journal} {Physical Review
  B}\ }\textbf {\bibinfo {volume} {105}},\ \bibinfo {pages} {125118} (\bibinfo
  {year} {2022})}\BibitemShut {NoStop}%
\bibitem [{\citenamefont {Maciejko}\ and\ \citenamefont
  {Rayan}(2022)}]{maciejko2022automorphic}%
  \BibitemOpen
  \bibfield  {author} {\bibinfo {author} {\bibfnamefont {J.}~\bibnamefont
  {Maciejko}}\ and\ \bibinfo {author} {\bibfnamefont {S.}~\bibnamefont
  {Rayan}},\ }\href@noop {} {\bibfield  {journal} {\bibinfo  {journal}
  {Proceedings of the National Academy of Sciences}\ }\textbf {\bibinfo
  {volume} {119}},\ \bibinfo {pages} {e2116869119} (\bibinfo {year}
  {2022})}\BibitemShut {NoStop}%
\bibitem [{\citenamefont {Urwyler}\ \emph {et~al.}(2022)\citenamefont
  {Urwyler}, \citenamefont {Lenggenhager}, \citenamefont {Boettcher},
  \citenamefont {Thomale}, \citenamefont {Neupert},\ and\ \citenamefont
  {Bzdu{\v{s}}ek}}]{urwyler2022hyperbolic}%
  \BibitemOpen
  \bibfield  {author} {\bibinfo {author} {\bibfnamefont {D.~M.}\ \bibnamefont
  {Urwyler}}, \bibinfo {author} {\bibfnamefont {P.~M.}\ \bibnamefont
  {Lenggenhager}}, \bibinfo {author} {\bibfnamefont {I.}~\bibnamefont
  {Boettcher}}, \bibinfo {author} {\bibfnamefont {R.}~\bibnamefont {Thomale}},
  \bibinfo {author} {\bibfnamefont {T.}~\bibnamefont {Neupert}},\ and\ \bibinfo
  {author} {\bibfnamefont {T.}~\bibnamefont {Bzdu{\v{s}}ek}},\ }\href@noop {}
  {\bibfield  {journal} {\bibinfo  {journal} {Physical Review Letters}\
  }\textbf {\bibinfo {volume} {129}},\ \bibinfo {pages} {246402} (\bibinfo
  {year} {2022})}\BibitemShut {NoStop}%
\bibitem [{\citenamefont {Zhang}\ \emph {et~al.}(2023)\citenamefont {Zhang},
  \citenamefont {Di}, \citenamefont {Zheng}, \citenamefont {Sun},\ and\
  \citenamefont {Zhang}}]{zhang2023hyperbolic}%
  \BibitemOpen
  \bibfield  {author} {\bibinfo {author} {\bibfnamefont {W.}~\bibnamefont
  {Zhang}}, \bibinfo {author} {\bibfnamefont {F.}~\bibnamefont {Di}}, \bibinfo
  {author} {\bibfnamefont {X.}~\bibnamefont {Zheng}}, \bibinfo {author}
  {\bibfnamefont {H.}~\bibnamefont {Sun}},\ and\ \bibinfo {author}
  {\bibfnamefont {X.}~\bibnamefont {Zhang}},\ }\href@noop {} {\bibfield
  {journal} {\bibinfo  {journal} {Nature Communications}\ }\textbf {\bibinfo
  {volume} {14}},\ \bibinfo {pages} {1083} (\bibinfo {year}
  {2023})}\BibitemShut {NoStop}%
\bibitem [{\citenamefont {Sun}\ \emph {et~al.}(2024)\citenamefont {Sun},
  \citenamefont {Chen}, \citenamefont {Bzdusek},\ and\ \citenamefont
  {Maciejko}}]{sun2024topological}%
  \BibitemOpen
  \bibfield  {author} {\bibinfo {author} {\bibfnamefont {C.}~\bibnamefont
  {Sun}}, \bibinfo {author} {\bibfnamefont {A.}~\bibnamefont {Chen}}, \bibinfo
  {author} {\bibfnamefont {T.}~\bibnamefont {Bzdusek}},\ and\ \bibinfo {author}
  {\bibfnamefont {J.}~\bibnamefont {Maciejko}},\ }\href@noop {} {\bibfield
  {journal} {\bibinfo  {journal} {SciPost Physics}\ }\textbf {\bibinfo {volume}
  {17}},\ \bibinfo {pages} {124} (\bibinfo {year} {2024})}\BibitemShut
  {NoStop}%
\bibitem [{\citenamefont {Guan}\ \emph {et~al.}(2025)\citenamefont {Guan},
  \citenamefont {Qi}, \citenamefont {Zhou}, \citenamefont {He},\ and\
  \citenamefont {Wang}}]{guan2025topological}%
  \BibitemOpen
  \bibfield  {author} {\bibinfo {author} {\bibfnamefont {D.-H.}\ \bibnamefont
  {Guan}}, \bibinfo {author} {\bibfnamefont {L.}~\bibnamefont {Qi}}, \bibinfo
  {author} {\bibfnamefont {Y.}~\bibnamefont {Zhou}}, \bibinfo {author}
  {\bibfnamefont {A.-L.}\ \bibnamefont {He}},\ and\ \bibinfo {author}
  {\bibfnamefont {Y.-F.}\ \bibnamefont {Wang}},\ }\href@noop {} {\bibfield
  {journal} {\bibinfo  {journal} {Physical Review B}\ }\textbf {\bibinfo
  {volume} {111}},\ \bibinfo {pages} {165144} (\bibinfo {year}
  {2025})}\BibitemShut {NoStop}%
\bibitem [{\citenamefont {Yuan}\ \emph {et~al.}(2024)\citenamefont {Yuan},
  \citenamefont {Zhang}, \citenamefont {Pei},\ and\ \citenamefont
  {Zhang}}]{yuan2024hyperbolic}%
  \BibitemOpen
  \bibfield  {author} {\bibinfo {author} {\bibfnamefont {H.}~\bibnamefont
  {Yuan}}, \bibinfo {author} {\bibfnamefont {W.}~\bibnamefont {Zhang}},
  \bibinfo {author} {\bibfnamefont {Q.}~\bibnamefont {Pei}},\ and\ \bibinfo
  {author} {\bibfnamefont {X.}~\bibnamefont {Zhang}},\ }\href@noop {}
  {\bibfield  {journal} {\bibinfo  {journal} {Physical Review B}\ }\textbf
  {\bibinfo {volume} {109}},\ \bibinfo {pages} {L041109} (\bibinfo {year}
  {2024})}\BibitemShut {NoStop}%
\bibitem [{\citenamefont {Lenggenhager}\ \emph {et~al.}(2023)\citenamefont
  {Lenggenhager}, \citenamefont {Maciejko},\ and\ \citenamefont
  {Bzdu{\v{s}}ek}}]{lenggenhager2023non}%
  \BibitemOpen
  \bibfield  {author} {\bibinfo {author} {\bibfnamefont {P.~M.}\ \bibnamefont
  {Lenggenhager}}, \bibinfo {author} {\bibfnamefont {J.}~\bibnamefont
  {Maciejko}},\ and\ \bibinfo {author} {\bibfnamefont {T.}~\bibnamefont
  {Bzdu{\v{s}}ek}},\ }\href@noop {} {\bibfield  {journal} {\bibinfo  {journal}
  {Physical Review Letters}\ }\textbf {\bibinfo {volume} {131}},\ \bibinfo
  {pages} {226401} (\bibinfo {year} {2023})}\BibitemShut {NoStop}%
\bibitem [{\citenamefont {Lux}\ and\ \citenamefont
  {Prodan}(2023)}]{lux2023converging}%
  \BibitemOpen
  \bibfield  {author} {\bibinfo {author} {\bibfnamefont {F.~R.}\ \bibnamefont
  {Lux}}\ and\ \bibinfo {author} {\bibfnamefont {E.}~\bibnamefont {Prodan}},\
  }\href@noop {} {\bibfield  {journal} {\bibinfo  {journal} {Physical Review
  Letters}\ }\textbf {\bibinfo {volume} {131}},\ \bibinfo {pages} {176603}
  (\bibinfo {year} {2023})}\BibitemShut {NoStop}%
\bibitem [{\citenamefont {Mosseri}\ and\ \citenamefont
  {Vidal}(2023)}]{mosseri2023density}%
  \BibitemOpen
  \bibfield  {author} {\bibinfo {author} {\bibfnamefont {R.}~\bibnamefont
  {Mosseri}}\ and\ \bibinfo {author} {\bibfnamefont {J.}~\bibnamefont
  {Vidal}},\ }\href@noop {} {\bibfield  {journal} {\bibinfo  {journal}
  {Physical Review B}\ }\textbf {\bibinfo {volume} {108}},\ \bibinfo {pages}
  {035154} (\bibinfo {year} {2023})}\BibitemShut {NoStop}%
\bibitem [{\citenamefont {El-Batanouny}\ and\ \citenamefont
  {Wooten}(2008)}]{el2008symmetry}%
  \BibitemOpen
  \bibfield  {author} {\bibinfo {author} {\bibfnamefont {M.}~\bibnamefont
  {El-Batanouny}}\ and\ \bibinfo {author} {\bibfnamefont {F.}~\bibnamefont
  {Wooten}},\ }\href@noop {} {\emph {\bibinfo {title} {Symmetry and condensed
  matter physics: a computational approach}}}\ (\bibinfo  {publisher}
  {Cambridge University Press},\ \bibinfo {year} {2008})\BibitemShut {NoStop}%
\bibitem [{\citenamefont {Hu}\ \emph {et~al.}(2023)\citenamefont {Hu},
  \citenamefont {Sun}, \citenamefont {Zeng}, \citenamefont {Ma}, \citenamefont
  {Dai}, \citenamefont {Yang}, \citenamefont {Zhang},\ and\ \citenamefont
  {Li}}]{hu2023source}%
  \BibitemOpen
  \bibfield  {author} {\bibinfo {author} {\bibfnamefont {C.}~\bibnamefont
  {Hu}}, \bibinfo {author} {\bibfnamefont {T.}~\bibnamefont {Sun}}, \bibinfo
  {author} {\bibfnamefont {Y.}~\bibnamefont {Zeng}}, \bibinfo {author}
  {\bibfnamefont {W.}~\bibnamefont {Ma}}, \bibinfo {author} {\bibfnamefont
  {Z.}~\bibnamefont {Dai}}, \bibinfo {author} {\bibfnamefont {X.}~\bibnamefont
  {Yang}}, \bibinfo {author} {\bibfnamefont {X.}~\bibnamefont {Zhang}},\ and\
  \bibinfo {author} {\bibfnamefont {P.}~\bibnamefont {Li}},\ }\href@noop {}
  {\bibfield  {journal} {\bibinfo  {journal} {Elight}\ }\textbf {\bibinfo
  {volume} {3}},\ \bibinfo {pages} {14} (\bibinfo {year} {2023})}\BibitemShut
  {NoStop}%
\bibitem [{\citenamefont {Li}\ \emph {et~al.}(2022)\citenamefont {Li},
  \citenamefont {Zhang}, \citenamefont {Budich},\ and\ \citenamefont
  {Trauzettel}}]{li2022transition}%
  \BibitemOpen
  \bibfield  {author} {\bibinfo {author} {\bibfnamefont {C.-A.}\ \bibnamefont
  {Li}}, \bibinfo {author} {\bibfnamefont {S.-B.}\ \bibnamefont {Zhang}},
  \bibinfo {author} {\bibfnamefont {J.~C.}\ \bibnamefont {Budich}},\ and\
  \bibinfo {author} {\bibfnamefont {B.}~\bibnamefont {Trauzettel}},\
  }\href@noop {} {\bibfield  {journal} {\bibinfo  {journal} {Physical Review
  B}\ }\textbf {\bibinfo {volume} {106}},\ \bibinfo {pages} {L081410} (\bibinfo
  {year} {2022})}\BibitemShut {NoStop}%
\bibitem [{\citenamefont {Koga}\ \emph {et~al.}(2021)\citenamefont {Koga},
  \citenamefont {Murakami},\ and\ \citenamefont {Nasu}}]{koga2021majorana}%
  \BibitemOpen
  \bibfield  {author} {\bibinfo {author} {\bibfnamefont {A.}~\bibnamefont
  {Koga}}, \bibinfo {author} {\bibfnamefont {Y.}~\bibnamefont {Murakami}},\
  and\ \bibinfo {author} {\bibfnamefont {J.}~\bibnamefont {Nasu}},\ }\href@noop
  {} {\bibfield  {journal} {\bibinfo  {journal} {Physical Review B}\ }\textbf
  {\bibinfo {volume} {103}},\ \bibinfo {pages} {214421} (\bibinfo {year}
  {2021})}\BibitemShut {NoStop}%
\bibitem [{\citenamefont {Wen}(2002)}]{wen2002quantum}%
  \BibitemOpen
  \bibfield  {author} {\bibinfo {author} {\bibfnamefont {X.-G.}\ \bibnamefont
  {Wen}},\ }\href@noop {} {\bibfield  {journal} {\bibinfo  {journal} {Physical
  Review B}\ }\textbf {\bibinfo {volume} {65}},\ \bibinfo {pages} {165113}
  (\bibinfo {year} {2002})}\BibitemShut {NoStop}%
\bibitem [{\citenamefont {Bieri}\ \emph {et~al.}(2016)\citenamefont {Bieri},
  \citenamefont {Lhuillier},\ and\ \citenamefont
  {Messio}}]{bieri2016projective}%
  \BibitemOpen
  \bibfield  {author} {\bibinfo {author} {\bibfnamefont {S.}~\bibnamefont
  {Bieri}}, \bibinfo {author} {\bibfnamefont {C.}~\bibnamefont {Lhuillier}},\
  and\ \bibinfo {author} {\bibfnamefont {L.}~\bibnamefont {Messio}},\
  }\href@noop {} {\bibfield  {journal} {\bibinfo  {journal} {Physical Review
  B}\ }\textbf {\bibinfo {volume} {93}},\ \bibinfo {pages} {094437} (\bibinfo
  {year} {2016})}\BibitemShut {NoStop}%
\bibitem [{\citenamefont {Chen}\ \emph
  {et~al.}(2023{\natexlab{a}})\citenamefont {Chen}, \citenamefont {Zhang},
  \citenamefont {Yang},\ and\ \citenamefont {Zhao}}]{chen2023classification}%
  \BibitemOpen
  \bibfield  {author} {\bibinfo {author} {\bibfnamefont {Z.}~\bibnamefont
  {Chen}}, \bibinfo {author} {\bibfnamefont {Z.}~\bibnamefont {Zhang}},
  \bibinfo {author} {\bibfnamefont {S.~A.}\ \bibnamefont {Yang}},\ and\
  \bibinfo {author} {\bibfnamefont {Y.}~\bibnamefont {Zhao}},\ }\href@noop {}
  {\bibfield  {journal} {\bibinfo  {journal} {Nature Communications}\ }\textbf
  {\bibinfo {volume} {14}},\ \bibinfo {pages} {743} (\bibinfo {year}
  {2023}{\natexlab{a}})}\BibitemShut {NoStop}%
\bibitem [{\citenamefont {Dusel}\ \emph {et~al.}(2025)\citenamefont {Dusel},
  \citenamefont {Hofmann}, \citenamefont {Maity}, \citenamefont {Mosseri},
  \citenamefont {Vidal}, \citenamefont {Iqbal}, \citenamefont {Greiter},\ and\
  \citenamefont {Thomale}}]{dusel2025chiral}%
  \BibitemOpen
  \bibfield  {author} {\bibinfo {author} {\bibfnamefont {F.}~\bibnamefont
  {Dusel}}, \bibinfo {author} {\bibfnamefont {T.}~\bibnamefont {Hofmann}},
  \bibinfo {author} {\bibfnamefont {A.}~\bibnamefont {Maity}}, \bibinfo
  {author} {\bibfnamefont {R.}~\bibnamefont {Mosseri}}, \bibinfo {author}
  {\bibfnamefont {J.}~\bibnamefont {Vidal}}, \bibinfo {author} {\bibfnamefont
  {Y.}~\bibnamefont {Iqbal}}, \bibinfo {author} {\bibfnamefont
  {M.}~\bibnamefont {Greiter}},\ and\ \bibinfo {author} {\bibfnamefont
  {R.}~\bibnamefont {Thomale}},\ }\href@noop {} {\bibfield  {journal} {\bibinfo
   {journal} {Physical Review Letters}\ }\textbf {\bibinfo {volume} {134}},\
  \bibinfo {pages} {256604} (\bibinfo {year} {2025})}\BibitemShut {NoStop}%
\bibitem [{\citenamefont {Lenggenhager}\ \emph {et~al.}(2025)\citenamefont
  {Lenggenhager}, \citenamefont {Dey}, \citenamefont {Bzdu{\v{s}}ek},\ and\
  \citenamefont {Maciejko}}]{lenggenhager2025hyperbolic}%
  \BibitemOpen
  \bibfield  {author} {\bibinfo {author} {\bibfnamefont {P.~M.}\ \bibnamefont
  {Lenggenhager}}, \bibinfo {author} {\bibfnamefont {S.}~\bibnamefont {Dey}},
  \bibinfo {author} {\bibfnamefont {T.}~\bibnamefont {Bzdu{\v{s}}ek}},\ and\
  \bibinfo {author} {\bibfnamefont {J.}~\bibnamefont {Maciejko}},\ }\href@noop
  {} {\bibfield  {journal} {\bibinfo  {journal} {Physical Review Letters}\
  }\textbf {\bibinfo {volume} {135}},\ \bibinfo {pages} {076604} (\bibinfo
  {year} {2025})}\BibitemShut {NoStop}%
\bibitem [{\citenamefont {Ellis}(2025)}]{HAP}%
  \BibitemOpen
  \bibfield  {author} {\bibinfo {author} {\bibfnamefont {G.}~\bibnamefont
  {Ellis}},\ }\href@noop {} {\bibinfo {title} {{HAP}, {Homological Algebra
  Programming}, {V}ersion 1.70}},\ \bibinfo {howpublished} {\href
  {https://gap-packages.github.io/hap}
  {\texttt{https://gap\texttt{\symbol{45}}packages.github.io/}\discretionary
  {}{}{}\texttt{hap}}, keywords = {homology; cohomology; resolution; homotopy
  group; module of identities; CW complex; simplicial complex; cubical complex;
  permutahedral complex; knots; nonabelian tensor; nonabelian exterior;
  covering space}, printedkey = {Ell25}} (\bibinfo {year} {2025}),\ \bibinfo
  {note} {gAP package}\BibitemShut {NoStop}%
\bibitem [{\citenamefont {Rober}(2024)}]{LINS0.9}%
  \BibitemOpen
  \bibfield  {author} {\bibinfo {author} {\bibfnamefont {F.}~\bibnamefont
  {Rober}},\ }\href@noop {} {\bibinfo {title} {{LINS}, provides an algorithm
  for computing the normal subgroups of a finitely presented group up to some
  given index bound., {V}ersion 0.9}},\ \bibinfo {howpublished} {\href
  {https://gap-packages.github.io/LINS/}
  {\texttt{https://gap-packages.github.io/}\discretionary
  {}{}{}\texttt{LINS/}}} (\bibinfo {year} {2024}),\ \bibinfo {note} {gAP
  package}\BibitemShut {NoStop}%
\bibitem [{\citenamefont {Chen}\ \emph
  {et~al.}(2023{\natexlab{b}})\citenamefont {Chen}, \citenamefont {Brand},
  \citenamefont {Helbig}, \citenamefont {Hofmann}, \citenamefont {Imhof},
  \citenamefont {Fritzsche}, \citenamefont {Kie{\ss}ling}, \citenamefont
  {Stegmaier}, \citenamefont {Upreti}, \citenamefont {Neupert} \emph
  {et~al.}}]{chen2023hyperbolic}%
  \BibitemOpen
  \bibfield  {author} {\bibinfo {author} {\bibfnamefont {A.}~\bibnamefont
  {Chen}}, \bibinfo {author} {\bibfnamefont {H.}~\bibnamefont {Brand}},
  \bibinfo {author} {\bibfnamefont {T.}~\bibnamefont {Helbig}}, \bibinfo
  {author} {\bibfnamefont {T.}~\bibnamefont {Hofmann}}, \bibinfo {author}
  {\bibfnamefont {S.}~\bibnamefont {Imhof}}, \bibinfo {author} {\bibfnamefont
  {A.}~\bibnamefont {Fritzsche}}, \bibinfo {author} {\bibfnamefont
  {T.}~\bibnamefont {Kie{\ss}ling}}, \bibinfo {author} {\bibfnamefont
  {A.}~\bibnamefont {Stegmaier}}, \bibinfo {author} {\bibfnamefont {L.~K.}\
  \bibnamefont {Upreti}}, \bibinfo {author} {\bibfnamefont {T.}~\bibnamefont
  {Neupert}}, \emph {et~al.},\ }\href@noop {} {\bibfield  {journal} {\bibinfo
  {journal} {Nature Communications}\ }\textbf {\bibinfo {volume} {14}},\
  \bibinfo {pages} {622} (\bibinfo {year} {2023}{\natexlab{b}})}\BibitemShut
  {NoStop}%
\bibitem [{\citenamefont {Chen}\ \emph {et~al.}(2024)\citenamefont {Chen},
  \citenamefont {Zhang}, \citenamefont {Qin}, \citenamefont {Bossart},
  \citenamefont {Yang}, \citenamefont {Chen},\ and\ \citenamefont
  {Fleury}}]{chen2024anomalous}%
  \BibitemOpen
  \bibfield  {author} {\bibinfo {author} {\bibfnamefont {Q.}~\bibnamefont
  {Chen}}, \bibinfo {author} {\bibfnamefont {Z.}~\bibnamefont {Zhang}},
  \bibinfo {author} {\bibfnamefont {H.}~\bibnamefont {Qin}}, \bibinfo {author}
  {\bibfnamefont {A.}~\bibnamefont {Bossart}}, \bibinfo {author} {\bibfnamefont
  {Y.}~\bibnamefont {Yang}}, \bibinfo {author} {\bibfnamefont {H.}~\bibnamefont
  {Chen}},\ and\ \bibinfo {author} {\bibfnamefont {R.}~\bibnamefont {Fleury}},\
  }\href@noop {} {\bibfield  {journal} {\bibinfo  {journal} {Nature
  Communications}\ }\textbf {\bibinfo {volume} {15}},\ \bibinfo {pages} {2293}
  (\bibinfo {year} {2024})}\BibitemShut {NoStop}%
\bibitem [{\citenamefont {Chan}\ \emph {et~al.}(2024)\citenamefont {Chan},
  \citenamefont {Andreanov}, \citenamefont {Flach},\ and\ \citenamefont
  {Batrouni}}]{chan2024superconductivity}%
  \BibitemOpen
  \bibfield  {author} {\bibinfo {author} {\bibfnamefont {S.~M.}\ \bibnamefont
  {Chan}}, \bibinfo {author} {\bibfnamefont {A.}~\bibnamefont {Andreanov}},
  \bibinfo {author} {\bibfnamefont {S.}~\bibnamefont {Flach}},\ and\ \bibinfo
  {author} {\bibfnamefont {G.~G.}\ \bibnamefont {Batrouni}},\ }\href@noop {}
  {\bibfield  {journal} {\bibinfo  {journal} {Physical Review B}\ }\textbf
  {\bibinfo {volume} {109}},\ \bibinfo {pages} {075153} (\bibinfo {year}
  {2024})}\BibitemShut {NoStop}%
\bibitem [{\citenamefont {Ara}\ \emph {et~al.}(2025)\citenamefont {Ara},
  \citenamefont {Banerjee}, \citenamefont {Basu},\ and\ \citenamefont
  {Krishnan}}]{ara2025flat}%
  \BibitemOpen
  \bibfield  {author} {\bibinfo {author} {\bibfnamefont {N.}~\bibnamefont
  {Ara}}, \bibinfo {author} {\bibfnamefont {A.}~\bibnamefont {Banerjee}},
  \bibinfo {author} {\bibfnamefont {R.}~\bibnamefont {Basu}},\ and\ \bibinfo
  {author} {\bibfnamefont {B.}~\bibnamefont {Krishnan}},\ }\href@noop {}
  {\bibfield  {journal} {\bibinfo  {journal} {SciPost Physics}\ }\textbf
  {\bibinfo {volume} {19}},\ \bibinfo {pages} {046} (\bibinfo {year}
  {2025})}\BibitemShut {NoStop}%
\bibitem [{\citenamefont {Fukui}\ and\ \citenamefont
  {Motome}(2025)}]{fukui2025topological}%
  \BibitemOpen
  \bibfield  {author} {\bibinfo {author} {\bibfnamefont {K.}~\bibnamefont
  {Fukui}}\ and\ \bibinfo {author} {\bibfnamefont {Y.}~\bibnamefont {Motome}},\
  }\href@noop {} {\bibfield  {journal} {\bibinfo  {journal} {Physical Review
  B}\ }\textbf {\bibinfo {volume} {112}},\ \bibinfo {pages} {144411} (\bibinfo
  {year} {2025})}\BibitemShut {NoStop}%
\end{thebibliography}%

\end{document}